\documentclass[traditabstract,longauth]{aa} 
\usepackage{graphicx}
\usepackage{natbib}
\usepackage{amsmath}
\usepackage{mathtools} 
\usepackage{fixltx2e} 
\usepackage{booktabs} 
\usepackage{rotating,multirow} 
\usepackage[UKenglish]{isodate}
\cleanlookdateon 
\def\deg{$^\circ$} 
\def\arcsec{$''$}
\def\muas{$\mu$as}
\def\muaspyr{$\mu$as/yr} 
\def\kms{km/s}
\def\mmag{mmag}

\def\teff{$T_{\rm eff}$}
\def\logg{$\log g$}
\def\feh{[Fe/H]}
\def\mh{[M/H]}
\def\a0{$A_{\rm 0}$}

\def\av{$A_{\rm V}$}
\def\atge{$A_{\rm TGE}$}
\def\relext{$R_{\rm 0}$}

\def\aabun{[$\alpha$/Fe]}
\def\vsini{$v\sin i$}

\def\gbp{$G_{\rm BP}$}
\def\grp{$G_{\rm RP}$}
\def\grvs{$G_{\rm RVS}$}
\def\cat{\ion{Ca}{ii}}

\def\ltsim{\ifmmode\stackrel{<}{_{\sim}}\else$\stackrel{<}{_{\sim}}$\fi}

\begin{document}

\title{The Gaia astrophysical parameters inference system (Apsis)}
\subtitle{Pre-launch description}
\titlerunning{Astrophysical parameters from Gaia}
\author{C.A.L.~Bailer-Jones\inst{1}\thanks{Email: calj@mpia.de} \and 
R.~Andrae\inst{1} \and
B.~Arcay\inst{2} \and
T.~Astraatmadja\inst{1} \and
I.~Bellas-Velidis\inst{3} \and
A.~Berihuete\inst{4} \and 
A.~Bijaoui\inst{5} \and
C.~Carri\'on\inst{6} \and
C.~Dafonte\inst{2} \and
Y.~Damerdji\inst{7,8} \and
A.~Dapergolas\inst{3} \and
P.~de Laverny\inst{5} \and
L.~Delchambre\inst{7} \and
P.~Drazinos\inst{9} \and
R.~Drimmel\inst{10} \and
Y.~Fr\'emat\inst{11} \and
D.~Fustes\inst{2} \and
M. Garc\'{i}a-Torres\inst{12} \and
C.~Gu\'ed\'e\inst{13,14} \and
U.~Heiter\inst{15} \and
A.-M.~Janotto\inst{16} \and
A.~Karampelas\inst{9} \and
D.-W.~Kim\inst{1} \and
J.~Knude\inst{17} \and
I.~Kolka\inst{18} \and
E.~Kontizas\inst{3} \and
M.~Kontizas\inst{9} \and
A.J.~Korn\inst{15} \and
A.C.~Lanzafame\inst{19,20} \and
Y.~Lebreton\inst{13,14} \and
H.~Lindstr{\o}m\inst{17,21} \and
C.~Liu\inst{1} \and
E.~Livanou\inst{9} \and
A.~Lobel\inst{11} \and
M.~Manteiga\inst{2} \and
C.~Martayan\inst{22} \and
Ch.~Ordenovic\inst{5} \and
B.~Pichon\inst{5} \and
A.~Recio-Blanco\inst{5} \and
B.~Rocca-Volmerange\inst{23,24} \and
L.M.~Sarro\inst{6} \and
K.~Smith\inst{1} \and
R.~Sordo\inst{25} \and
C.~Soubiran\inst{26} \and
J.~Surdej\inst{7} \and
F.~Th\'evenin\inst{5} \and
P.~Tsalmantza\inst{1} \and
A.~Vallenari\inst{25} \and
J.~Zorec\inst{23}
}
\institute{
Max Planck Institute for Astronomy, K\"onigstuhl 17, 69117 Heidelberg, Germany \and
Fac.\ Inform\'atica, Universidade da Coru\~na, Campus de Elvi\~na 15071, A Coru\~na, Spain \and
Institute for Astronomy, Astrophysics, Space Applications \& Remote Sensing, National Observatory of Athens, PO Box 20048, 11810 Athens, Greece \and
Dpt. Statistics and Operations Research, University of C\'adiz, Campus Universitario R\'{\i}o San Pedro s/n. 11510 Puerto Real, C\'adiz, Spain \and
Laboratoire Lagrange (UMR7293), Universit\'e de Nice Sophia Antipolis, CNRS, Observatoire de la C\^ote d'Azur, BP 4229, 06304 Nice, France \and
Dpt. de Inteligencia Artificial, UNED, Juan del Rosal, 16, 28040 Madrid, Spain \and
Institut d'Astrophysique et de G\'eophysique, Universit\'e de Li\`ege, Allée du 6 Ao\^ut, 17, B5C, 4000 Sart Tilman, Belgium \and
Centre de Recherche en Astronomie, Astrophysique et G\'eophysique, Route de l'Observatoire Bp 63 Bouzareah, DZ-16340 Algiers, Algeria \and
Department of Astrophysics, Astronomy \& Mechanics, Faculty of Physics, University of Athens, 15783 Athens, Greece \and
Osservatorio Astrofisico di Torino, Istituto Nazionale di Astrofisica (INAF), Pino Torinese, Italy \and
Royal Observatory of Belgium, 3 avenue circulaire, B1180 Brussels, Belgium \and
Area of Languages and Computer Systems, Pablo de Olavide University, Seville, Spain \and
Observatoire de Paris, GEPI, CNRS UMR 8111, 92195, Meudon, France \and
Institut de Physique de Rennes, Universit\'e de Rennes 1, CNRS UMR 6251, F-35042 Rennes, France \and
Division of Astronomy and Space Physics, Department of Physics and Astronomy, Uppsala University, Box 516, SE-751\,20 Uppsala, Sweden \and
Centre National d'Etudes Spatiales, 18 av Edouard Belin, 31401 Toulouse, France \and
Niels Bohr Institute, Copenhagen University, Copenhagen, Denmark \and
Tartu Observatory, 61602 Toravere, Estonia \and
Dipartimento di Fisica e Astronomia, Sezione Astrofisica, Universit\'a di Catania, via S. Sofia 78, Catania, Italy \and
INAF - Osservatorio Astrofisico di Catania, via S. Sofia 78, Catania, Italy \and
CSC Danmark A/S, Retortvej 8, DK-2500 Valby, Denmark \and
European Southern Observatory, Alonso de Cordova 3107, Vitacura, Santiago de Chile, Chile \and
Institut d’Astrophysique de Paris, UMR 7095 CNRS – Universit\'e Pierre \& Marie Curie, 98bis boulevard Arago, 75014 Paris, France \and
Universit\'e de Paris-Sud XI, IAS, 91405, Orsay Cedex, France \and
INAF - Osservatorio Astronomico di Padova, Padova, Italy \and
LAB UMR 5804, Univ.\ Bordeaux  - CNRS, Floirac, France
}
\date{submitted 23 July 2013; revised 8 September 2013; accepted 9 September 2013}

\abstract{ The Gaia satellite will survey the entire celestial sphere down to 20th magnitude, obtaining astrometry, photometry, and low resolution spectrophotometry on one billion astronomical sources, plus radial velocities for over one hundred million stars.  Its main objective is to take a census of the stellar content of our Galaxy, with the goal of revealing its formation and evolution. Gaia's unique feature is the measurement of parallaxes and proper motions with hitherto unparalleled accuracy for many objects.  As a survey, the physical properties of most of these objects are unknown. Here we describe the data analysis system put together by the Gaia consortium to classify these objects and to infer their astrophysical properties using the satellite's data.  This system covers single stars, (unresolved) binary stars, quasars, and galaxies, all covering a wide parameter space. Multiple methods are used for many types of stars, producing multiple results for the end user according to different models and assumptions. Prior to its application to real Gaia data the accuracy of these methods cannot be assessed definitively.  But as an example of the current performance, we can attain internal accuracies (RMS residuals) on F,G,K,M dwarfs and giants at $G$\,=\,15 ($V$\,=\,15--17) for a wide range of metallicites and interstellar extinctions of around 100\,K in effective temperature (\teff), 0.1\,mag in extinction (\a0), 0.2\,dex in metallicity (\feh), and 0.25\,dex in surface gravity (\logg).  The accuracy is a strong function of the parameters themselves, varying by a factor of more than two up or down over this parameter range.  After its launch in
November 2013, Gaia will nominally observe for five years, during which the system we describe will continue to evolve in light of experience with the real data. }
\keywords{galaxies: fundamental parameters -- methods: data analysis -– methods: statistical -– stars: fundamental parameters -– surveys: Gaia} 
\maketitle

\section{Introduction}\label{sect:introduction}

The ESA Gaia satellite will provide the most extensive astrometric survey of our Galaxy to date. Its primary mission is to measure the positions, parallaxes, and proper motions for essentially all objects in the sky between visual (G-band) magnitudes 6 and 20, some 10$^9$ stars and several million galaxies and quasars. By revealing the three-dimensional distribution and space motions of a statistically significant sample of stars across the whole Galaxy, Gaia will enable a fundamentally new type of exploration of the structure, formation and evolution of our Galaxy. 
Furthermore, this exquisite astrometry -- parallax uncertainties as low as 10\,\muas\ -- will promote major advances in our knowledge and understanding of stellar structure, open clusters, binary stars and exoplanets, lead to discoveries of near-earth asteroids and provide tests of general relativity
(e.g.\  \citealp{2001A&A...369..339P}, \citealp{2005ESASP.576.....T}, \citealp{2008IAUS..248..217L},
\citealp{2008A&A...482..699C}, \citealp{2009IAUS..254..475B}, \citealp{2010IAUS..261..306M}, \citealp{2012P&SS...73....5T}).

To achieve these goals, astrophysical information on the astrometrically measured sources is indispensable. For this reason Gaia is equipped with two low resolution prism spectrophotometers, which together provide the spectral energy distribution of all targets from 330 to 1050\,nm. Data from these spectrophotometers (named BP and RP for ``blue photometer'' and ``red photometer'') will be used to classify sources and to determine their astrophysical parameters (APs), such as stellar metallicities, line-of-sight extinctions, and the redshifts of quasars. The spectrophotometry is also required to correct the astrometry for colour-dependent shifts of the image centroids. Spectra from the higher resolution radial velocity spectrograph (RVS, 847--871\,nm) on board will provide further information for estimating APs as well as some individual abundances for the brighter stars.

Gaia scans the sky continuously, building up data on sources over the course of its five year mission. Its scanning strategy, plus the need for a sophisticated self-calibration of the astrometry, demands an elaborate data processing procedure. It involves numerous interdependent operations on the data, including photometric processing, epoch cross-matching, spectral reconstruction, CCD calibration, attitude modelling, astrometric parameter determination, flux calibration, astrophysical parameter estimation, and variability analysis, to name just a few. These tasks are the responsibility of a large academic consortium, the {\em Data Processing and Analysis Consortium} (DPAC), comprising over 400 members in 20 countries. The DPAC comprises nine coordination units (CUs), each dealing with a different aspect of the data processing.

One of these CUs, CU8 ``Astrophysical Parameters'', is responsible for classifying and estimating the astrophysical parameters of the Gaia sources. In this article we describe the data processing system developed to achieve this goal. This system, called {\em Apsis}, comprises a number of modules, each of which will be described here.

The Gaia data processing is organized into a series of consecutive cycles centered around a versioned main data base (MDB). At the beginning of each operation cycle, the various processes read the data they need from the MDB. At the end of the cycle, the results are written to a new version of the MDB which, together with new data from the satellite, forms the MDB for the next processing cycle. In this way, all of the Gaia data will be sequentially processed until, at some version of the MDB several cycles after the end of observations, all data have received all necessary treatment and the final catalogue can be produced. Some suitably processed and calibrated data will be siphoned off during the processing into intermediate data releases, expected to start about two years after launch \citep{LL:TJP-011, LL:XL-033}.  
For more details of the overall processing methodology see \citet{LL:FM-030}, \citet{2007ASPC..376...99O}, and \citet{2008IAUS..248..224M}.

We continue our presentation of the Gaia astrophysical processing system in section~\ref{sect:gaiadata} by looking more closely at the data Gaia will provide. Section~\ref{sect:apsis} gives an overview of Apsis: the guiding concepts behind it, its component modules and how they interact, and how it will be used during the mission. Section~\ref{sect:training} describes the data we have used for model training and testing. In section~\ref{sect:algorithms} we describe each of the modules and give some impression of the results which can be expected. More details on several of these can be found in published or soon-to-be published articles. In section~\ref{sect:validation} we outline how we plan to validate and calibrate the system once we get the Gaia data, and how we might improve the algorithms during the mission. We wrap up in section~\ref{sect:outlook}.  More information on the Gaia mission, the data processing and planned data releases, as well as some of the DPAC technical notes cited, can be obtained from \url{http://www.rssd.esa.int/Gaia}.

\section{Gaia observations and data}\label{sect:gaiadata}

An overview of the Gaia instruments, their properties and expected performance can be found in \citet{2012Ap&SS.341...31D} and at \url{http://tinyurl.com/GaiaPerformance}.
Here we summarize some essential features relevant to our description of Apsis.

\subsection{Overview and observation strategy}

Gaia observes continuously, its two telescopes -- which share a focal plane -- scanning a great circle on the sky as the satellite rotates, once every six hours. The satellite simultaneously precesses with a period of 63 days. The combined result of these motions is that the entire sky is observed after 183 days.  Each source is therefore observed a number of times over the course of the mission.  These multiple observations, made at different points on Gaia's orbit around the Sun, are the basis for the astrometric analysis \citep{2012A&A...538A..78L}.

As the satellite rotates, a source sweeps across a large focal plane mosaic of CCDs. These are read out synchronously with the source motion (``time-delayed integration'', TDI).  Over the first 0.7\deg\ 
of the focal plane scan, the source is observed in unfiltered light -- the $G$-band -- for the purpose of the astrometry.
Further along the light is dispersed by two prisms to produce the BP/RP spectrophotometry.  At the trailing edge the light is dispersed by a spectrograph to deliver the RVS spectra.

Although Gaia observes the entire sky, not all CCD pixels are transmitted to the ground. Gaia selects, in real-time, windows around point sources brighter than $G$\,=\,20. The profile of the $G$-band, spanning 330--1050\,nm, is defined by the mirror and CCD response \citep{2010A&A...523A..48J}.\footnote{The $G-V$ colours for B1V, G2V, and M6V stars are $-0.01$, $-0.18$ and $-2.27$\,mag respectively, so the Gaia limiting magnitude of $G$\,=\,20 corresponds to $V$\,=\,20--22 depending on the spectral type.}  The source detection is near-diffraction limited to about 0.1\arcsec\ (the primary mirrors have dimensions 1.45m$\times$0.5m).  While most of the $10^9$ sources we expect Gaia to observe will be stars, a few million will be quasars and galaxies with point-like cores, and asteroids. \citet{2012A&A...543A.100R} give predictions of the number, distribution and types of sources which will be observed.

Each source will be observed between 40 and 250 times, depending primarily on its ecliptic latitude.  BP/RP spectrophotometry are nominally obtained for all sources at every epoch. These are combined during the processing and calibration into a single BP/RP spectrum for each source. RVS spectra are obtained at fewer epochs due to the focal plane architecture, and these are also combined.

The accuracy of AP estimation depends strongly on the spectral signal-to-noise ratio (SNR) which, for a given number of epochs, is primarily a function of the source's G magnitude.  In the rest of this article we will consider a single BP/RP spectrum to be a combination of 70 observation epochs, which is the sky-averaged number of epochs per source (accounting also for various sources of epoch loss).  For RVS it is 40 epochs.
We refer to such combined spectra as ``end-of-mission'' spectra.  All results in this article were obtained using (simulated) end-of-mission spectra.

\subsection{Spectrophotometry (BP/RP) and spectroscopy (RVS)}\label{sect:specdata}

\begin{figure*}
\begin{center}
\includegraphics[width=1.0\textwidth]{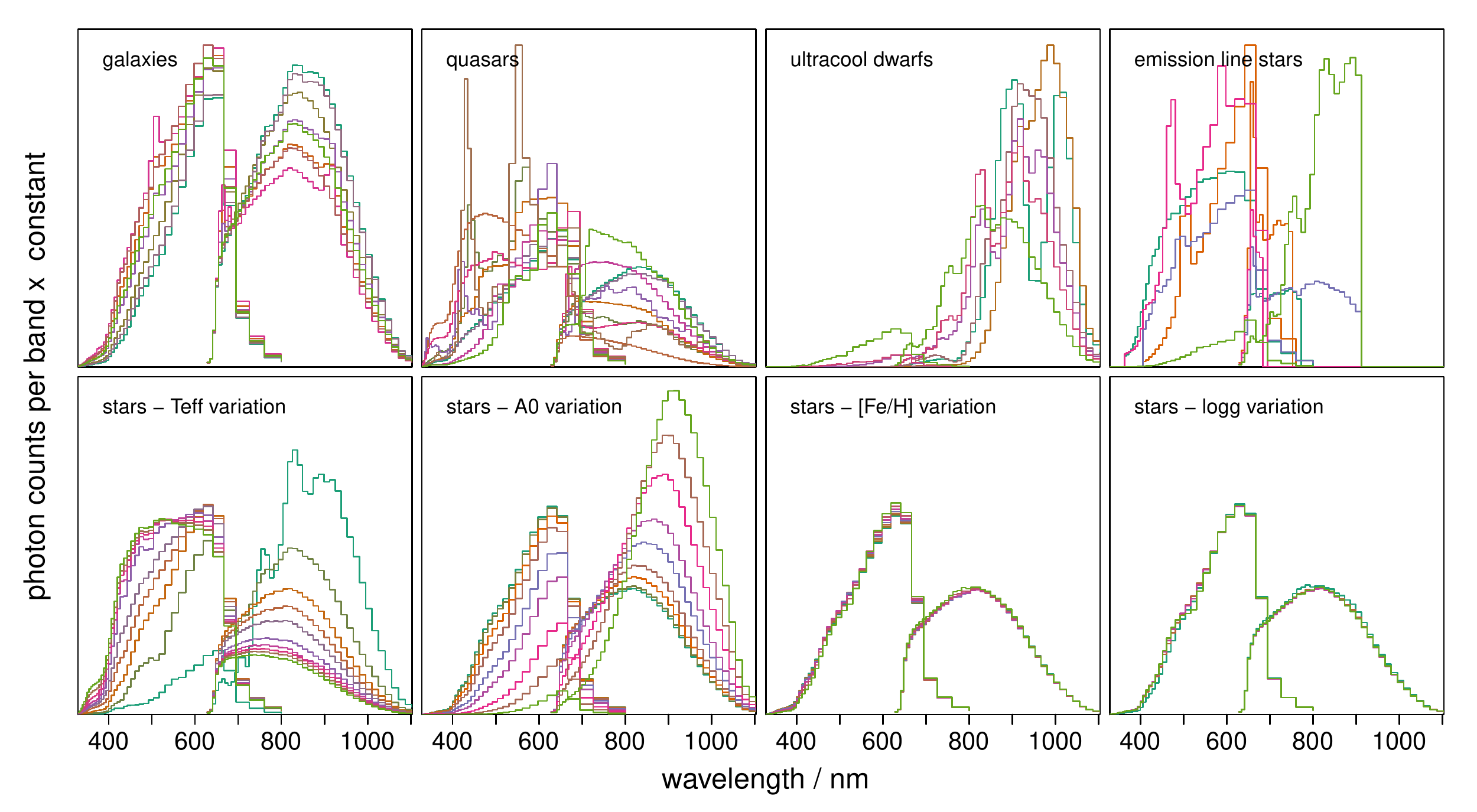}
\caption[]{Example BP/RP spectrophotometry.
The spectra have been normalized to have the same number of photon counts over the spectral bands plotted. (This does not yield the same area under each spectrum as plotted -- against wavelength -- due to the nonlinear dispersion.)
Except for the emission line stars, all spectra are noise-free synthetic spectra.
Several examples of each type of object are shown in each panel (the line colours are arbitrary).
The galaxies are for a range of types, all with zero redshift and zero Galactic extinction.
The quasar spectra cover a range of emission line strengths, continuum slopes and redshifts.
In the ultra cool dwarf panel six spectra are shown with \teff\ ranging from 500--3000\,K in steps of 500\,K for \logg\,=\,5\,dex.
The top-right panel shows five emission line sources: Herbig Ae, PNe, T Tauri, WN4/WCE, dMe. The lack of red flux for these sources is a result of the input spectra used not spanning the full BP/RP wavelength range.
The bottom row shows normal stars, in which just one parameter varies in each panel. From left to right these are: \teff\,$\in\{3000, 4000, 5000, 6000, 7000, 8500, 10000, 12000, 15000, 20000\}$\,K; \a0\,$\in\{0.0, 0.1, 0.5, 1.0, 2.0, 3.0, 5.0, 7.0, 10.0\}$\,mag; \feh\,$\in\{-2.5, -1.5, -0.5, +0.5\}$\,dex; \logg\,$\in\{0, 2.5, 4, 5.5\}$\,dex.
The other parameters are held constant as appropriate at \teff\,=\,5000\,K, \a0\,=\,0\,mag, \feh\,=\,0\,dex, \logg\,=\,4.0\,dex.
A common photon count scale is used in all the panels of the bottom row. The cooler/redder stars are those with increasingly more flux in the red part of the spectrum in the two lower left panels.
\label{fig:bprpspectra}}
\end{center}
\end{figure*}

\begin{figure}
\begin{center}
\includegraphics[width=0.4\textwidth]{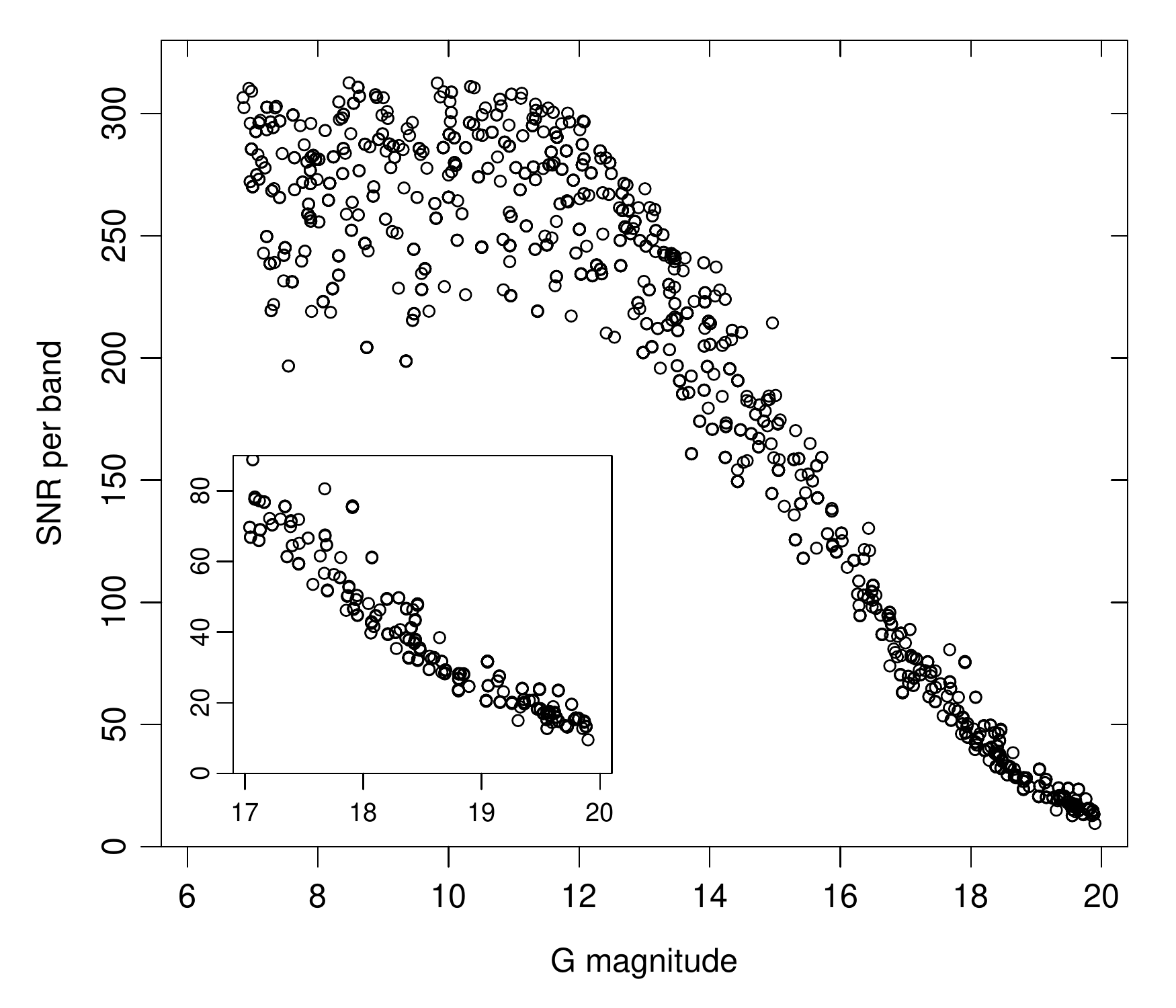}
\caption[]{Variation of the SNR per band in end-of-mission BP/RP for a set of 2000 stars covering the HR diagram. 
The inset is a zoom of the fainter magnitudes. 
Discontinuities occur at several brighter magnitudes (barely visible here) on account of the use of TDI gates to limit the integration time for bright stars in order to avoid saturation of the CCDs.
The SNR for each spectrum is the mean over the bands plotted in Figure~\ref{fig:bprpspectra}.
In addition to the formal noise model errors, an additional error of 0.3\% in the flux has been added in quadrature to accommodate calibration errors.
\label{fig:snrbprpvsmag}
}
\end{center}
\end{figure}

BP and RP spectra are read out of the CCDs with 60 wavelength samples (or ``bands''; they can be thought of as narrow overlapping filters).  BP spans 330--680\,nm with a resolution (=$\lambda / \Delta\lambda$) varying from 85 to 13, and RP spans 640--1050\,nm with a resolution of 26 to 17 \citep{2012Ap&SS.341...31D}. ($\Delta\lambda$ is defined as the 76\% energy width of the line spread function.)
The resolution is considerably lower than what one would like for AP estimation, but limitations are set by numerous factors.\footnote{The Gaia consortium optimized a multi-band photometric system for Gaia, described by \citet{2006MNRAS.367..290J}, but due to mission constraints this was not adopted.}  
The upstream processing can in principle deliver spectra of higher resolution by a factor of a few for all sources, because the multiple epoch spectra are offset by fractions of a sample. Such ``oversampled spectra'' are not used in the present work.
Examples of star, galaxy and quasar spectra are shown in Figure~\ref{fig:bprpspectra}.
Low SNR bands at the edges of both BP and RP have been omitted (approximately 8 bands from each end of both).
Apsis will use BP/RP to classify all Gaia sources and to estimate APs down to the Gaia magnitude limit, although some ``weaker'' APs, such as \logg, will be poorly estimated at $G$\,=\,20.  

The expected variation of SNR of BP/RP with magnitude is shown in Figure~\ref{fig:snrbprpvsmag}. This plot includes a calibration error corresponding to 0.3\% of the flux, an estimate based both on past experience and our current understanding of the impact of systematic errors.
This has not yet been included in the synthetic spectra used to train and test most Apsis modules, because it is difficult to estimate its magnitude in advance.  Ignoring this calibration error increases the SNR at $G$\,=\,15 from about 175 to around 225, and for $G>17$ the difference in SNR is 10\% or less, so most of our results are unaffected by this.  Without systematic errors the SNR would extend to 1000--2000 for $G<12$.

\begin{figure}
\begin{center}
\includegraphics[width=0.50\textwidth]{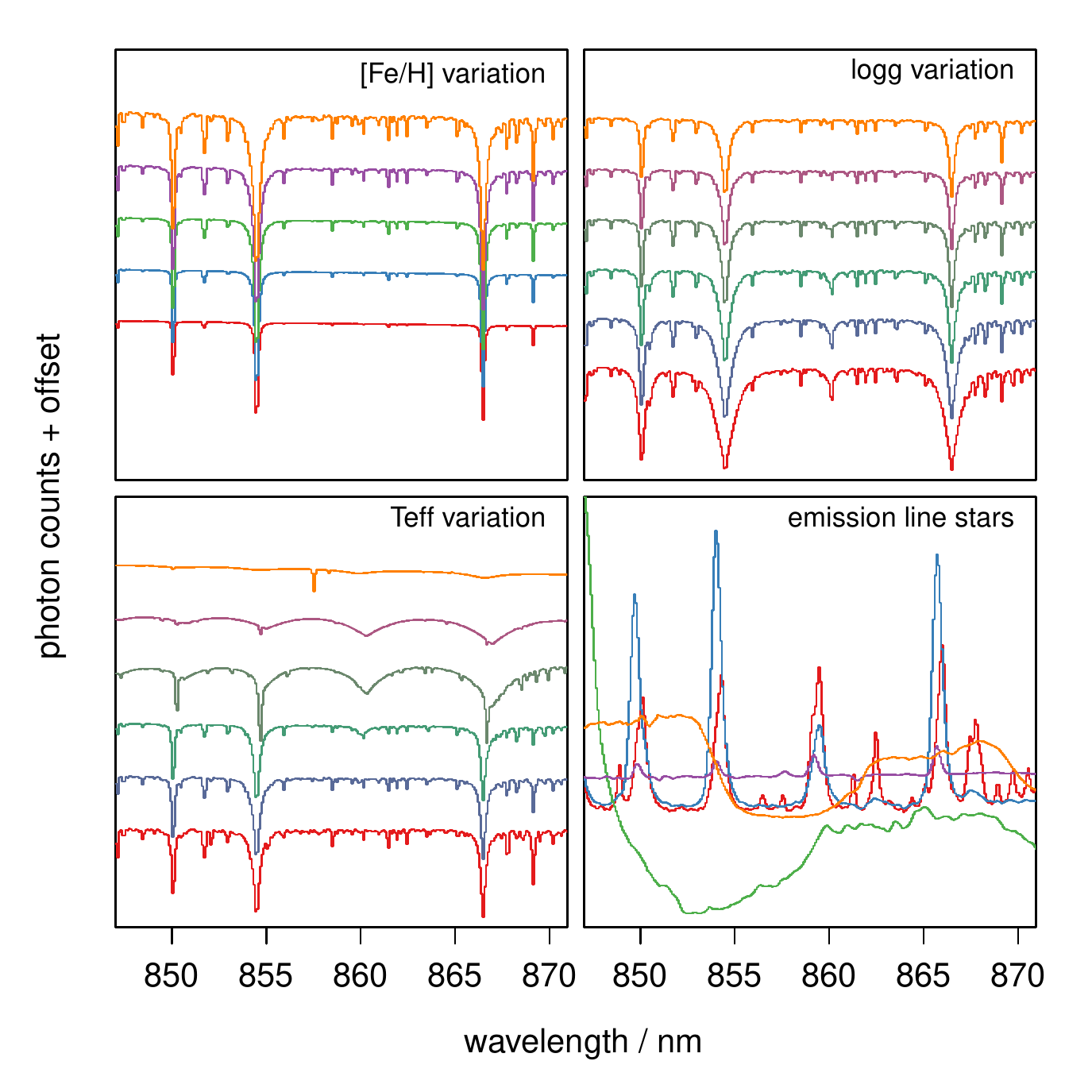}
\caption[]{Example RVS spectra.
Each spectrum is noise free and at \grvs\,=\,12 (i.e.\ with RVS in low resolution mode).  The line colours are arbitrary.
Three of the panels show the variation of one of the APs with the other two held constant, the constant values being
\teff=5500\,K, \feh=0\,dex, \logg=4.0\,dex. \aabun=0\,dex in all cases.
In these cases the spectra in each panel have been offset vertically for clarity.
The AP ranges (increasing from bottom to top in each panel) are:
\feh=$-2.5$ to $0.0$ in steps of 0.5\,dex; \logg=$0$ to $5$ in steps of 1\,dex;
\teff\,$\in\{4500, 5500, 6500, 8250, 14000, 40000\}$\,K.
The bottom right panel shows examples of five emission line stars (here the offset is zero). They are, from bottom to top around the feature at 859\,nm: nova; WC star; O6f star; Be star; B[e] star.
\label{fig:rvsspectra}}
\end{center}
\end{figure}

\begin{figure}
\begin{center}
\includegraphics[width=0.40\textwidth]{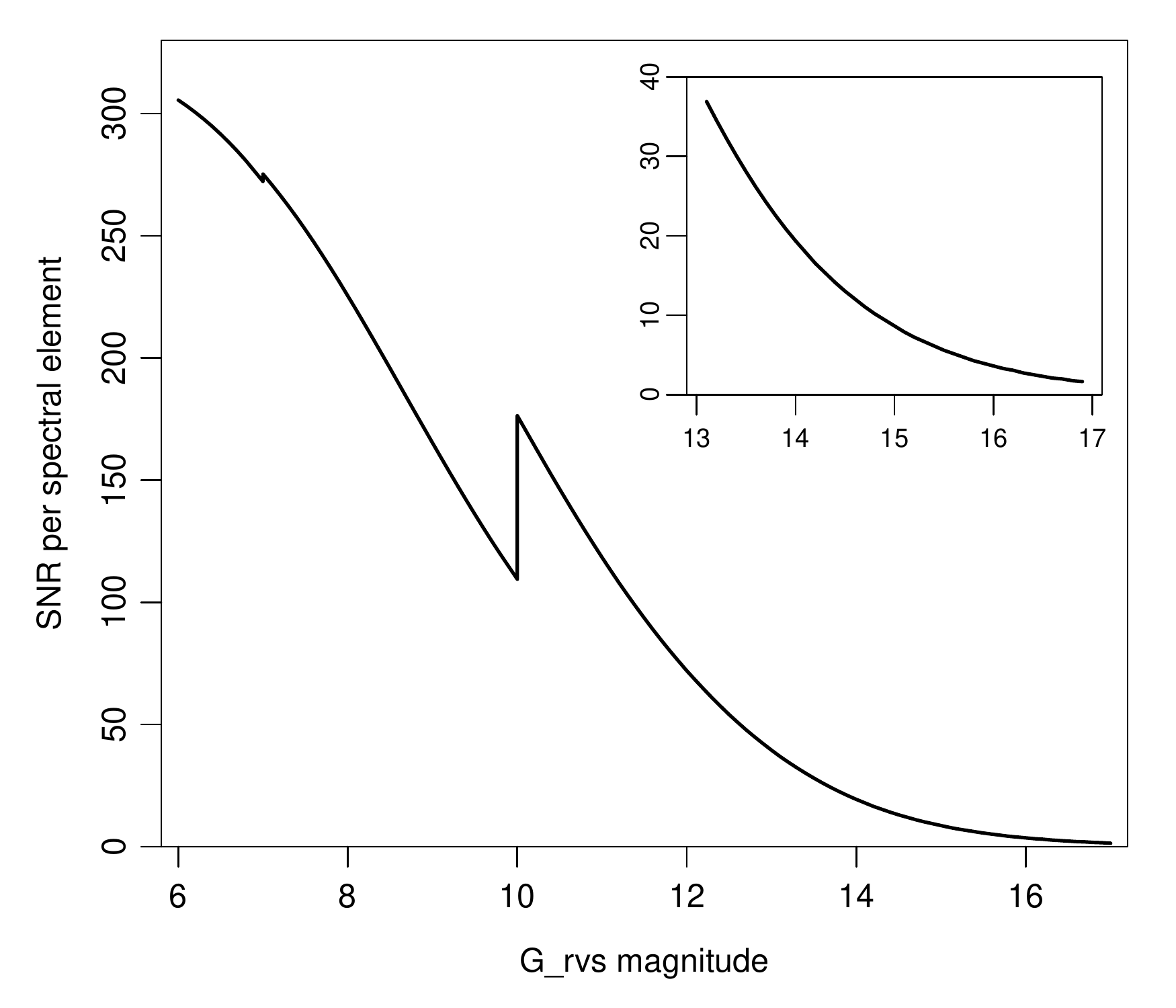}
\caption[]{Variation of the SNR per spectral element in end-of-mission RVS spectra with \grvs. The inset is a zoom of the fainter magnitudes. 
The discontinuity at \grvs\,=\,10 is due to the on-chip binning of the spectrum in the dispersion direction for fainter stars, and that at \grvs\,=\,7 is due to on-chip binning perpendicular to the dispersion direction for fainter stars.
In addition to the formal noise model errors, an additional error of 0.3\% in the flux has been added in quadrature to accommodate calibration and normalization errors.
\label{fig:snrrvsvsmag}
}
\end{center}
\end{figure}

The radial velocity spectrograph \citep{2004MNRAS.354.1223K,2011EAS....45..181C} records spectra from 847 to 871\,nm (the \cat\ triplet region) at a resolution of 11\,200; the figures given here reflect the manufactured instruments (T.\ Prusti September 2012, private communication).
For SNR reasons, RVS does not extend to the $G$\,=\,20 limit of the other instruments, but will be limited to 
about \grvs\,=\,17, so is expected to deliver useable spectra for of order 200 million stars.\footnote{\grvs\ is the photometric band formed by integrating the RVS spectrum, and in terms of magnitudes \grvs\,$\simeq I_{\rm C}$.}
Spectra fainter than \grvs\,=\,10 are binned on-chip by a factor of three in the dispersion direction in order to improve the SNR, at the cost of a lower spectral resolution. 
The main purpose of RVS is to measure radial velocities -- the sixth component of the phase space. The radial velocity precision for most stars ranges from 1--15\,\kms, depending strongly on both colour and magnitude
(\citealp{2011EAS....45..189K}, \citealp{2012Ap&SS.341...31D}).
Apsis uses RVS data both for general stellar parameter estimation down to about
\grvs\,=\,14.5 (of order 35 million stars), and for characterizing specific types, such as emission line objects.

Examples of the RVS spectra are shown in Figure~\ref{fig:rvsspectra}.  The typical variation of the SNR with \grvs\ is shown in Figure~\ref{fig:snrrvsvsmag}.  This plot includes a 0.3\% error assumed to arise from imperfect calibration and normalization. This was not included in the synthetic libraries used to train and test Apsis modules, although it decreases the SNR by no more than 15\% for \grvs$>10$. It must be appreciated, however, that obtaining useable RVS spectra at the faint end depends critically on how well charge transfer inefficiency (CTI) effects in the CCDs can be modelled \citep{2012MNRAS.419.2995P}.

The extraction, combination and calibration of both BP/RP and RVS spectra are complicated tasks which will not be discussed here. They are the responsibility of the coordination units CU5 (for BP/RP) and CU6 (for RVS) in DPAC, and are discussed in various technical notes (e.g.\ \citealp{2011EAS....45..149J,2011EAS....45..189K,LL:FDA-025}).
Apsis works with ``internally calibrated'' BP/RP and RVS spectra, by which we mean they are all on a common flux scale (and various CCD phenomena have been removed), but the instrumental profile and dispersion function have not been removed.  The library spectra which form the basis of training our classification modules are projected into this data space using an instrument simulator (see section~\ref{sect:training}).

\subsection{Photometry and astrometry}

The sources' G-band magnitudes are measured to a precision of 1--3\mmag, limited by calibration errors even at $G$\,=\,20.
The data processing will also produce integrated photometry for BP, RP and RVS, with magnitudes referred to as \gbp, \grp, and \grvs\ respectively. For more details of these passbands including transformations between them and to non-Gaia passbands, see \citet{2010A&A...523A..48J}. These bands are used in Apsis primarily to assess (together with the number of observation epochs) the SNR of the spectra.

The astrometry is used in Apsis to help distinguish between Galactic and extragalactic objects, and parallaxes are also used in a few modules to aid stellar AP estimation.
The astrometric accuracy is a function mostly of SNR and thus G magnitude. At $G$=15, 18.5, and 20 the sky-averaged parallax accuracy is 25--26\,\muas, 137--145\,\muas, and 328--347\,\muas\ respectively, the ranges reflecting the colour dependence across early B to late M stellar spectral types (slightly better for earlier type stars).\footnote{On account of the large width of the G-band, the accuracies at constant V-band magnitude are quite different, e.g.\
ranging from 26\,\muas\ for early B types to 9\,\muas\ for late M types at $V$=15.}
For $6<G<14$ the accuracy is 7--17\muas, although the performance at the bright limit will depend on the actual TDI gate scheme used to avoid saturating the bright stars.\footnote{The 5\,000 or so stars brighter than $G$\,=\,6 will saturate in the focal plane, but may yield useful measurements if we can calibrate their diffraction spikes.}
The proper motion accuracies in \muaspyr\ are about 0.5 times the size of the quoted parallax accuracies.

How the parallax accuracy converts to distance accuracy depends on the parallax itself.  For example, an unreddened K1 giant at 5\,kpc would have an apparent magnitude of $G=14.0$ and a distance accuracy of 9\%. 
A G3 dwarf at 2\,kpc has $G=16.5$ and a distance accuracy of 8\%.  When combining these accuracies with a model for the Galaxy, we expect the number of stars with distance determinations better than 0.1\%, 1\% and 10\% to be of order $10^5$, $10^7$, and $10^8$ (respectively).

\section{The astrophysical parameters inference system (Apsis)}\label{sect:apsis}

\subsection{Principles}

The goal of Apsis is to classify and to estimate astrophysical parameters for the Gaia sources using the Gaia data. These APs will be part of the publicly released Gaia catalogue. They will also be used internally in the data processing, for example to help the template-based extraction of the RVS spectra and the identification of quasars used to fix the astrometric reference frame.

Our guiding principle for Apsis is to provide reasonably accurate estimates for a broad class of objects covering a large fraction of the catalogue, rather than to treat some specific types of objects exhaustively. To achieve this, Apsis consists of a number of modules with different functions.

The paradigm which underlies most of the Apsis modules is supervised learning. This means that the classes or parameters of objects are determined according to the similarity of the data to a set of templates for which the parameters are already known, so-called ``labelled'' data.  How this comparison is done -- in particular, how we interpolate between the templates and how we use the data -- is an important attribute distinguishing between the various machine learning (or pattern recognition) algorithms available.  Our choices are based on their accuracy, utility and speed.  The term ``training'' is used to describe the process by which the algorithm is fit to (learns from) the template data.  For the most part we have, to date, used libraries of synthetic spectra as the basis for our training data, although we also use some semi-empirical libraries. These libraries and the construction of the training and testing data using a Gaia instrument simulator are described in section~\ref{sect:training}. Later, actual Gaia observations will be used to calibrate the synthetic spectral grids (see section~\ref{sect:validation}).

\subsection{Architecture}

\begin{figure}
\begin{center}
\includegraphics[width=0.5\textwidth]{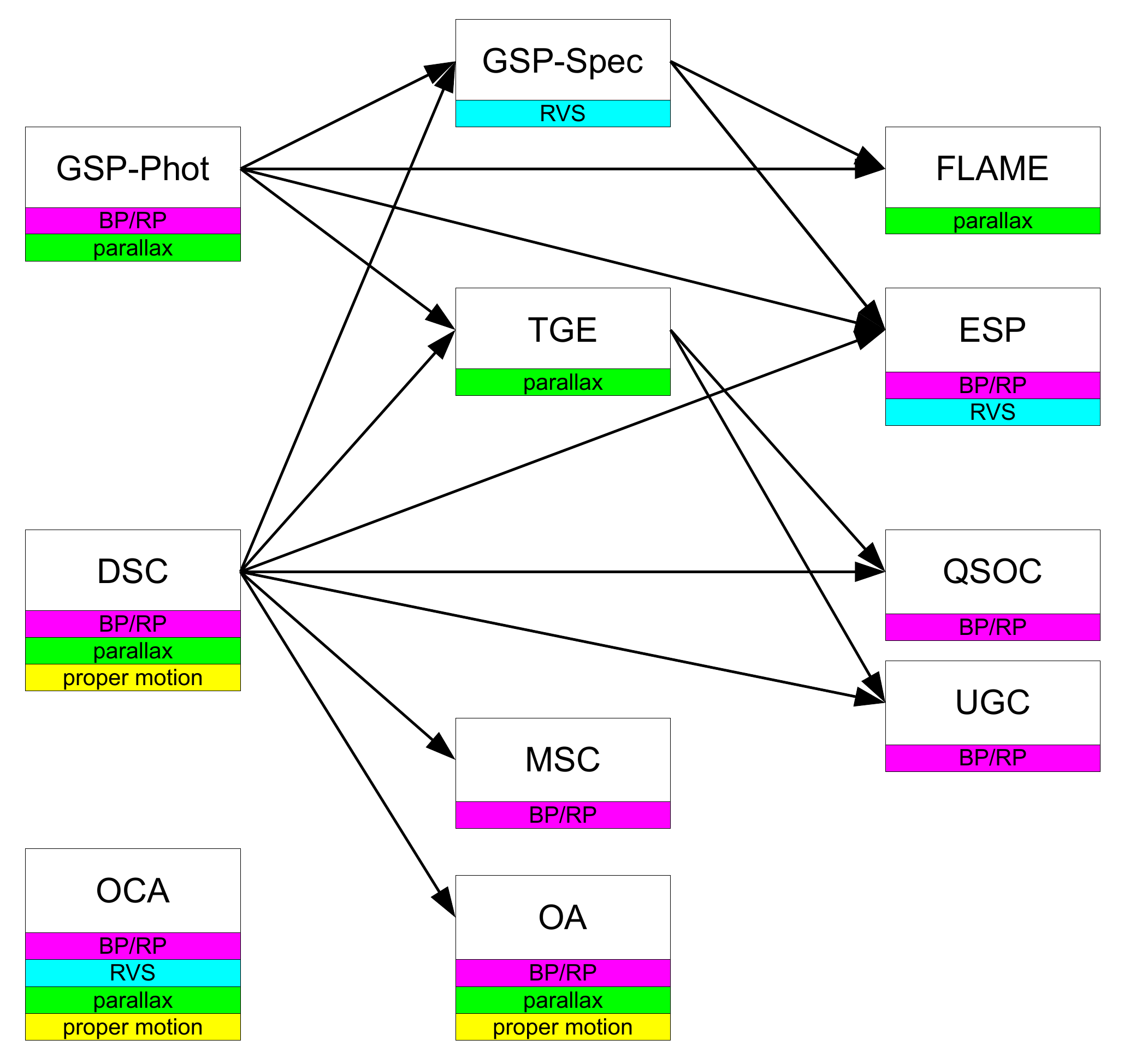}
\caption[]{Component modules in Apsis and their interdependency. The module names are defined in Table~\ref{tab:spnames}. The arrows indicate a dependency on the output of the preceding module. The coloured bars underneath each module indicate which data it uses. Most of the modules additionally use the photometry and some also the Galactic coordinates.
\label{fig:apsisarchitecture}}
\end{center}
\end{figure}

Each of the modules in Apsis is described separately in section~\ref{sect:algorithms}. Here we give an overview and describe their connectivity, which is summarized in Figure~\ref{fig:apsisarchitecture}. The acronyms are defined in Table~\ref{tab:spnames}.

\begin{table}
\begin{center}
\caption{Apsis modules \label{tab:spnames}}
\footnotesize{
\begin{tabular}{ll}
\toprule
Acronym & name \\
\midrule
DSC & Discrete Source Classifier \\
ESP & Extended Stellar Parametrizer: \\
\hspace*{2em}-CS & ESP -- Cool Stars \\
\hspace*{2em}-ELS & ESP -- Emission Line Stars \\
\hspace*{2em}-HS & ESP -- Hot Stars \\
\hspace*{2em}-UCD & ESP -- Ultra Cool Dwarfs \\
FLAME & Final Luminosity Age and Mass Estimator \\
GSP-Phot & Generalized Stellar Parametrizer -- Photometry \\
GSP-Spec & Generalized Stellar Parametrizer -- Spectroscopy \\
MSC & Multiple Star Classifier \\
OA & Outlier Analysis \\
OCA & Object Clustering Algorithm \\
QSOC & Quasar Classifier \\
TGE & Total Galactic Extinction\\
UGC & Unresolved Galaxy Classifier \\
\bottomrule
\end{tabular}
}
\end{center}
\end{table}

DSC performs a probabilistic classification into classes such as ``(single) star'', ``binary star'', ``quasar''. This is used by many of the other modules to select sources for processing.  GSP-Phot and GSP-Spec estimate stellar parameters using the BP/RP spectra (and parallaxes) and the RVS spectra respectively, whereby GSP-Phot also estimates the line-of-sight extinction to each star individually. Supporting these are a number of ``extended stellar parametrization'' modules, which operate on specific types of stars, their preliminary identification being taken from GSP-Phot and (if the stars are bright enough) GSP-Spec.  These are ESP-ELS, ESP-HS, ESP-CS, and ESP-UCD.  Although GSP-Phot is trained on a broad set of stars which includes all of these, these modules attempt to achieve more appropriate parameters estimates by making a more physically-motivated use of the data, and/or by using other stellar models. Using the outputs of GSP-Phot, FLAME uses isochrones to estimate stellar luminosities, masses and ages for certain types of stars. MSC attempts to estimate parameters of both components of systems suspected (by DSC) to be unresolved stellar binaries. QSOC and UGC estimate astrophysical parameters of quasars and (unresolved) galaxies, respectively.  TGE will use the line-of-sight extinction estimates from GSP-Phot of the most distant stars to build a two-dimensional map of the total Galactic extinction over the whole sky. This may also be used as an input to QSOC.

The two remaining Apsis modules use the concept of unsupervised learning. OCA works independently of all other modules by using clustering techniques to detect ``natural'' patterns in the data, primarily for novelty detection. 
OA does something similar on the objects classified as ``outliers'' by DSC. Its purpose is to identify whether some of these outliers are known objects which were not, or were not correctly, modelled in the training data. Results from this can be used to improve the models in the next processing cycle.

\subsection{Source selection}

Which sources are processed by which modules depends on (1) the availability of the necessary data; (2) the SNR of the data; (3) the outputs from other modules.

DSC operates on all sources which have BP/RP data, which is nominally all Gaia sources.  For each source, DSC assigns probabilities to a set of classes. This is the main output from DSC for the end-user. In addition, a single ``best'' class will be identified for each source. In principle this is just the class which receives the highest probability, but in practice this probability will also have to exceed some class-dependent threshold.  Some sources may not attain this, in which case they will be classified as ``unknown''.

GSP-Phot operates on all sources too. As more than 99\% of sources are expected to be stars, there is little loss of efficiency if GSP-Phot is simply applied to everything, regardless of the DSC class. The GSP-Phot APs for what are later chosen to be non-stars based on the DSC class probabilities can then simply be ignored.

GSP-Spec will operate on all stars identified by DSC which have RVS spectra with sufficiently high SNR (\grvs\,\ltsim\,15).  FLAME operates on a subset of sources which have APs of sufficient precision from GSP-Phot and/or GSP-Spec.  TGE selects a small fraction of distant stars assigned precise extinction estimates by GSP-Phot.  The remaining modules, specifically ESP-HS, ESP-CS, ESP-ELS, ESP-UCD, MSC, QSOC, and UGC, will only be applied to objects of ``their'' class, as determined by the DSC class probabilities. 

\subsection{Multiple parameter assignments}\label{sect:multipleAPs}

A consequence of our system design is that any given source may be assigned multiple sets of APs.  For example, a particular star could be assigned APs by GSP-Phot, ESP-CS, and GSP-Spec.  This is an inevitable consequence of a diverse approach to inference: the conclusions we draw depend not only on the data we measure, but also on the stellar models we adopt (as embodied in the training data) and other assumptions made. We can never know the true APs with 100\% confidence. All of these sets of APs will be reported in the MDB and the data releases, thus giving the end-user the freedom to choose among our models and assumptions.  For those users who would rather forgo this choice, we will also provide the ``best'' set of APs for each source.  We will establish the criteria for making this decision during the operations, based on experience with the data. GSP-Phot estimates APs for all stars, so there will always be a homogeneous set of stellar APs available.

The situation is actually more complex than this, because a few of the modules themselves comprise multiple algorithms, each providing separate estimates of the APs. One of the reasons for this is cross-checking: if two or more algorithms give similar results for the same source (and training data), our confidence in the results may be increased. A second reason is that different methods may make use of different data. For example, the Aeneas algorithm in GSP-Phot (section~\ref{sect:gspphot}) can operate with or without the parallax. The former is potentially more accurate, yet makes more assumptions, so we may be interested in both results. A third reason for using multiple algorithms is that the best performing algorithm may be computationally too expensive to run on all sources.

\subsection{Scope}

One of the principles we adopt in DPAC is that the Gaia catalogue will be based only on Gaia data for the individual sources. (Non-Gaia observations are used for validation and calibration; see section~\ref{sect:validation}.)  ``Better'' AP estimates could be obtained for some sources by including external data in the analysis, such as higher resolution spectra or infrared photometry.  The DPAC objective, however, is to produce a homogeneous Gaia catalogue by processing all sources in a consistent manner.
We hope that the community at large will extend our work by using the published data to make composite analyses where appropriate.

While Apsis tries to cover most types of objects, it does not include everything.  Asteroids are excluded, for example. They will be detected by Gaia primarily via their very large proper motions, so they will be classified by the CU charged with their detection (CU4).  Apsis presently ignores morphological information. Although Gaia only tracks point sources, two-dimensional images could be reconstructed using the multiple scans at different orientations over a source. This is planned by other CUs in DPAC, and could be useful for further galaxy characterization, for example. This may be introduced later into the data processing.

Apsis also does not take into account stellar variability. An entire CU in DPAC, CU7, is dedicated to classifying variable stars from, primarily, their G-band light curves. As Apsis works with combined epoch spectra, some types of variable source will receive spurious APs. During the course of the data processing we will investigate how and whether variability information can be introduced into our work.

\subsection{Software, hardware and operations}

The Apsis modules have been developed by various CU8 groups over the past years following a cyclic development process. They are written in the Java programming language according to DPAC-wide software engineering standards. The modules are integrated into a control system which deals with job allocation and data input/output.

As outlined in section~\ref{sect:introduction}, the Gaia data processing proceeds in cycles, centered around the MDB.  When Apsis first runs (several months into the mission), essentially all Gaia sources will have been observed at least once. In succeeding cycles, Apsis will run again on the same set of sources, but the data are the combination of more observation epochs, so will have higher SNR and improved calibrations. The first significant, calibrated results from Apsis should appear about 2.5 years into the mission, and will be made available in the subsequent intermediate data release.

Apsis will run on multicore computers at CU8's data processing centre hosted by CNES in Toulouse. The time Apsis needs for processing is likely to vary considerably during the ongoing development, but as of late 2012, the supervised modules (i.e.\ excluding OA and OCA) together required of order 15 GFLOP (1 GFLOP = $10^9$ floating point operations) for a single source. This is dominated by the Markov Chain Monte Carlo (MCMC) sampling performed by the Aeneas algorithm in GSP-Phot. 
A common CPU will today provide around 100 GFLOP per second, so processing all $10^9$ Gaia sources in this way would take 1740 days. The Apsis processing is trivial to parallelize, so running it on 100 CPU cores reduces this to 17 days.  However, the 15 GFLOPS figure neglects data input/output, which is likely to be a considerable fraction of the processing time. OCA and OA are also likely to add to this figure significantly.  On the other hand, CU8 will have more like 400 CPU cores available full time for its processing. Given that we need to process all Gaia sources in one operation cycle (duration of 6--12 months), these figures are acceptable even if we assume some intra-cycle reprocessing.

\section{Model training and testing}\label{sect:training}

Supervised classification methods are based on the comparison of observed data with a set of templates. These are used to train the models in some way.  For this purpose we may use either observed or synthetic templates, both of which have
their advantages and disadvantages.
Observed templates better represent the spectra one will actually encounter in the real data, but rarely cover the necessary parameter range with the required density, in particular not for a survey mission like Gaia. Synthetic templates allow us to characterize a wide parameter space, and also to model sources which are very rare or even which have not (yet) been observed.  Intrinsically free of observational noise and interstellar extinction, they allow us to freely add these effects in a controlled manner.  They are, however, simplifications of the complex physics and chemistry in real astrophysical sources, so they do not reproduce real spectra perfectly. This may be problematic for pattern recognition, so synthetic spectra will need calibration using the actual Gaia observations of known sources (see section~\ref{sect:validation}).\footnote{Of course, to estimate {\em physical} parameters we must, at some point, use {\em physical} models, so dependence on synthetic spectra cannot be eliminated entirely.}

The training data for the Apsis modules are based on a mixture of observed (actually ``semi-empirical'') and synthetic libraries for the main sources we expect to encounter. These are described below. Once the library spectra have been constructed, BP/RP and RVS spectra are artificially reddened, then simulated at the required G magnitude and with a SNR corresponding to end-of-mission spectra (see section~\ref{sect:gaiadata}) using the Gaia Object Generator (GOG, \citealp{2005ESASP.576..357L}).

\subsection{Stellar spectral libraries}

\begin{table*}
\caption{
Stellar libraries used to simulate BP/RP and RVS spectra. $N$ is the number of spectra in the library.
Ap/Bp are peculiar stars; UCD are ultracool dwarfs; WR are Wolf Rayet stars; WD are white dwarfs. 
\label{tab:speclibs}
}
\begin{center}
\begin{tabular}{l r r@{$-$}r c c r l}
\toprule
Name            &$N$&  \multicolumn{2}{c}{\teff\,/\,K} & \logg\,/\,dex  & \feh\,/\,dex   & Ref.$^\dagger$ & Notes    \\
\midrule
OB stars        & 1296 & 15\,000&55\,000  & 1.75$-$4.75 &0.0$-$0.6  & $1$   & \scriptsize TLUSTY code; NLTE, mass loss, $v_{\rm micro}$\\ 
Ap/Bp stars     &  36  & 7000&16000   &   4.0       &0.0          & $2$   & \scriptsize  LLmodels code, chemical peculiarities   \\ 
A stars         & 1450 & 6000&16\,000   & 2.5$-$4.5   &0.0          & $3$   & \scriptsize  LLmodels code, \aabun\,=\,0.0, +0.4   \\ 
MARCS           & 1792 &  2800&8000   &$-$0.5$-$5.5 &$-$5.0$-$1.0  & $4$   & \scriptsize Galactic enrichment law for \aabun\   \\ 
Phoenix         & 4575 & 3000&10\,000   &$-$0.5$-$5.5 &$-$2.5$-$0.5  & $5$   & \scriptsize $\Delta$\teff=100~K   \\ 
UCD             & 2560 &  400&4000    &$-$0.5$-$5.5 &$-$2.5$-$0.5  & $6$   & \scriptsize various dust models   \\ 
C stars MARCS    &  428 &  4000&8000   & 0.0$-$5.0   &$-$5.0$-$0.0  & $7$   & \scriptsize [C/Fe] depends on \feh\  \\ 
Be              &  174 &15\,000&25\,000   &     4.0     & 0.0         & $8$   & \scriptsize range of envelope to stellar radius ratios   \\ 
WR              &  43  &25\,000&51\,000   &  2.8$-$4.0  & 0.0         & $9$   & \scriptsize range of mass loss rates    \\ 
WD              & 187  &  6000&90\,000  &  7.0$-$9.0  &  0.0          & $10$  & \scriptsize WDA \& WDB     \\ 
MARCS NLTE      &  33  &  4000&6000   &  4.5$-$5.5  & 0.0         & $11$  & \scriptsize  NLTE line profiles  \\ 
MARCS RVS       &146\,394& 2800&8000    &$-$0.5$-$5.5 &$-$5.0$-$1.0  & $12$  & \scriptsize  variations in individual elements abundances   \\ 
3D models        &  13  &4500&6500     &  2.0$-$5.0   &$-$2.0$-$0.0  & $13$  & \scriptsize  StaggerCode models and Optim3D code \\ 
SDSS stars      & 50\,000 &  3750&10\,000  &  0.0$-$5.5  &$-$2.5$-$0.5  & $14$ & \scriptsize  semi-empirical library  \\
Emission line stars & 1620 &  &  & $-$ & $-$ & $15$ & \scriptsize semi-empirical library (see section~\ref{sect:espels}) \\
\bottomrule
\end{tabular}
\end{center}
$^\dagger$\scriptsize{References: 
1 \cite{2008RMxAC..33...50B}; 2 \cite{2008CoSka..38..419K}; 3 \cite{2004AA...428..993S};
4 \cite{2008AA...486..951G};  5 \cite{2005ESASP.576..565B}; 6 \cite{2001ApJ...556..357A};
7 Masseron, priv.\ comm.;  8, 9 \cite{2008sf2a.conf..499M};
10 \cite{2006AA...450..331C}; 11 Korn et al.\ 2009, priv.\ comm.; 12 Recio-Blanco et al.\ 2011, priv.\ comm.; 13 \cite{2011JPhCS.328a2012C}; 14 \cite{LL:PAT-008}; 15 \citet{LL:AJL-001}}
\end{table*}

The Gaia community has calculated large libraries of synthetic spectra with improved physics for many types of stars. 
We are able to cover a broad AP space with some redundancy between libraries.  Each library uses codes optimized for a given \teff\ range, or for a specific object type, and includes as appropriate the following phenomena: departures from local thermodynamic equilibrium (LTE); dust; mass loss; circumstellar envelopes; magnetic fields; variations of single element abundances; chemical peculiarities. The libraries are listed in Table~\ref{tab:speclibs} with a summary of their properties and AP space. Not all of these libraries are used in the results reported in section~\ref{sect:algorithms}. The synthetic stellar libraries are described in more detail in \citet{2010Ap&SS.328..331S, 2011JPhCS.328a2006S} together with details on their use in the Gaia context. 
The large synthetic grids for A, F, G, K, and M stars have been computed in LTE for both BP/RP and RVS.  For OB stars, non-LTE (NLTE) line formation has been taken into account. 

Synthetic spectra are of course not perfect. We cannot yet satisfactorily simulate some processes, such as emission line formation. To mitigate these drawbacks, observed spectra are included in the training dataset in the form of semi-empirical libraries.  These are observed spectra to which APs have been assigned using synthetic spectra, and for which the wavelength coverage has been extended (as necessary) using the best fitting synthetic spectrum. Semi-empirical libraries have been constructed for ``normal'' stars using SDSS \citet{LL:PAT-008}, and from other sources for emission line stars (\citealp{LL:AJL-001}; see section~\ref{sect:espels}).

Starting from the available synthetic and semi-empirical libraries, two types of data set are produced. The first one mirrors the AP space of the spectral libraries, and is regularly spaced in some APs. The second one involves interpolation on some of the APs (\teff, \logg, \feh) but with no extrapolation (and we do not combine different libraries). See \citet{2011JPhCS.328a2006S} for details on how the interpolation is done.  Both datasets are intended for training the AP estimation modules, while the interpolated one serves also for testing. In both cases extinction is applied using Cardelli's law \citep{1989ApJ...345..245C}, with a given set of extinction parameters. Extinction is represented using an extinction parameter, \a0, rather than the extinction in a particular band, as defined in section 2.2 of \citet{2011MNRAS.411..435B}.  The parameter \a0\ corresponds to \av\ in Cardelli’s formulation of the extinction law, but this new formulation is chosen to clarify that it is an extinction parameter, and not necessarily the extinction in the $V$ band, because the extinction (for broad bands) depends also on the spectral energy distribution of the source.

Mass, radius, age, and absolute magnitudes of the stars are derived using the Padova evolutionary models \citep{2008AA...484..815B}, resulting in a full description of stellar sources. These models cover a wide range of masses up to 100\,M$_\odot$ and metallicities for all evolutionary phases. Although this is not needed for most Apsis modules, it is required by the GSP-Phot module Aeneas when it is using parallax, in order to ensure consistency between parallax, apparent and absolute magnitude, and the stellar parameters in the training data set (see section 3.3 of \citealp{2012MNRAS.426.2463L} for a discussion).

\subsection{Galaxy spectral libraries}\label{sect:galspeclib}

For the classification of unresolved galaxies we have generated synthetic spectra of normal galaxies \citep{2007A&A...470..761T, 2009A&A...504.1071T, 2012A&A...538A..38K} using the galaxy evolution model P\'EGASE.2 \citep{1997A&A...326..950F,1999astro.ph.12179F}.  The objective was not just to obtain a set of typical synthetic spectra, but to have a broad enough sample which can predict the full variety of galaxies we expect to observe with Gaia. Four galaxy spectral types have been adopted: early, spiral, irregular, quenched star formation.  For each type, a restframe spectrum is characterized by four APs: the timescale of in-falling gas and three parameters which define the appropriate star formation law. The current library comprises 28\,885 synthetic galaxy spectra at zero redshift. These are then simulated at a range of redshifts from 0 to 0.2, and a range of values of \a0\ from 0 to 10\,mag in order to simulate extinction due to the interstellar medium of our Galaxy.

In addition, a semi-empirical library of 33\,670 galaxy spectra has been produced fitting SDSS spectra to the synthetic galaxy library \citep{2012A&A...537A..42T}.

\subsection{Quasar spectral libraries}

Two different libraries of synthetic spectra of quasars have been generated, one with regular AP space coverage (17\,325 spectra) and one with random AP space coverage (20\,000 spectra). Three QSO APs are sampled: redshift (from 0 to 5.5), slope of the continuum ($\alpha$ from -4 to +3), and equivalent width of the emission lines (EW from $10^1$ to $10^5$\,nm).  The libraries also sample a broad range of Galactic interstellar extinction, \a0\,=\,0--10\,mag.

A semi-empirical library of 70\,556 DR7 SDSS spectra of quasars has also been generated.  The majority of quasars in this library have redshift below 4, $\alpha$ from -1 to +1 and EW from 0 to 400\,nm (the distributions are very non-uniform).  These are not artificially reddened, but some will have experienced a small amount of real extinction.

\section{The Apsis modules}\label{sect:algorithms}

We now describe each of the modules in Apsis listed in Table~\ref{tab:spnames}
and summarized in Figure~\ref{fig:apsisarchitecture}. 

\subsection{Discrete Source Classifier (DSC)}\label{sect:dsc}

DSC performs the top-level classification of every Gaia source, assigning a probability to each of a number of classes.  These are currently: star, white dwarf, binary star, galaxy, quasar.  For this it uses three groups of input data: the BP/RP spectrophotometry; the proper motion and parallax; the position and the photometry in the $G$-band. Each group of input data is directed to a separate subclassifier (described below), each of which produces a vector of probabilities for the classes. The results from all the subclassifiers are combined into a single probability vector, and based on this a class label may be generated if the highest probability exceeds a certain threshold.  An additional module using the $G$-band light curve (time series) -- or rather metrics extracted from it -- is under development.  An earlier phase of the DSC development was presented in some detail by \cite{2008MNRAS.391.1838B}, who examine in particular the issue of trying to identify rare objects.

The photometric subclassifier works with the BP/RP spectra. The classification algorithm is the Support Vector Machine (SVM; \citealp{Vapnik:1995:NSL:211359, 1995MachineLearning.20.273C, bur98}), which is widely used for analysing high-dimensional astrophysical data (e.g.~\citealp{2010A&A...522A..88S}).  We use the implementation libSVM \citep{cha01}.  (We have also tested other machine learning algorithms, such as random forests, and find similar overall performance.)  A set of SVM models is trained, each at a different $G$ magnitude range, and the observed BP/RP spectrum passed to the one appropriate to its measured magnitude. (This is done because SVMs work best when the training and test data have similar noise levels.) Each model contains two layers, the first trained on an astrophysically meaningful distribution of common objects, the second trained on a broad distribution of AP space and intended in particular to classify rare objects. For each layer, a front-end outlier detector identifies sources that do not resemble the training data sufficiently closely. Only objects rejected by the first layer are passed to the second layer for classification.  
Flags are set to indicate outliers detected by each layer. These will be studied by the OA module (see section~\ref{sect:oa}).

The astrometric subclassifier uses the parallaxes and proper motions to help distinguish between Galactic and extragalactic objects.
This uses a three-dimensional Gaussian mixture model trained on noise-free, simulated astrometry for Galactic and extragalactic objects. This model is convolved with the estimated uncertainties in the proper motion and parallax for each source, and a probability of the source being Galactic or extragalactic is calculated.

The position--magnitude subclassifier gives a probability of the source being Galactic or extragalactic based on the source's position and brightness. This reflects our broad knowledge of the overall relative frequency of Galactic and extragalactic objects and how they vary as a function of magnitude and Galactic coordinates. For example, if we knew only that a source had $G$=14 and were at low Galactic latitude, we would think it more likely to be Galactic than extragalactic.  In the absence of more data, we should fall back on this prior information. This subclassifier quantifies this using a simple lookup table based on a simple universe model.  For very informative spectra, this subclassifier would have little influence on the final probabilities.

\begin{table}
\begin{center}
\caption{Example of the DSC classification performance shown as a confusion matrix for sources with magnitudes in the range G=6.8--20. The rows indicate the true classes (the spectral libraries), the columns the DSC assigned class. Each cell gives the percentage of objects classified from each true class to each DSC class. The dashes indicate exactly zero. 
 \label{tab:dscResults}
}
\begin{tabular}{lrrrrr}
\toprule
        & \multicolumn{5}{c}{Output class} \\
Library       & Star  & WD  & Binary & Quasar & Galaxy \\
\midrule
Phoenix     &  91.9 &  $-$ &  7.1   &  $-$   &  1.0   \\
Phoenix--\relext\  & 89.9  &  3.0 &  7.1   &  $-$   & $-$ \\
A stars          &  79.9 &  $-$ & 20.0   &  $-$   &  0.1   \\
OB stars         &  95.3 &  0.6 &  4.1   &  $-$   &  $-$   \\
WD               &  17.4 & 79.1 &  3.5   &  $-$   &  $-$ \\  
UCDs          &  97.3 &  $-$ & 1.0    &  $1.7$   & $-$ \\
Binary stars &  18.3 &  $-$ & 81.7   &  $-$   &  $-$ \\ 
SDSS stars       &  94.1 &  $-$ &  5.9   &  $-$   &  $-$ \\
SDSS quasars     &   5.9 &  3.0 &  0.1   & 78.3   & 12.7 \\   
SDSS galaxies    &   2.0 &  $-$ &  0.5   &  $-$   & 97.5 \\ 
\bottomrule
\end{tabular}
\end{center}
\end{table}

DSC is trained on numerous data sets built from almost all of the spectral libraries described in section~\ref{sect:training}, including blended spectra of different types of objects (e.g.\ optical stellar binaries). A selection of the results on independent test sets constructed from these libraries is shown in Table~\ref{tab:dscResults}.  Phoenix--\relext\ is the Phoenix library but now also showing a large variation in the second extinction parameter \relext.  (More detailed results from an earlier version of the software can be found in \citealp{LL:KS-019}.)  These results combine the outputs from all three subclassifiers, and is for Galactic objects with magnitudes ranging from $G=$\,6.8--20 and quasars and galaxies from $G=$\,14--20 (uniform distributions) in both the training and test sets.  The synthetic spectra include the 0.3\% calibration error mentioned in section~\ref{sect:specdata}. The SDSS stars, quasars and galaxies are the semi-empirical libraries.  The performance for stars and galaxies is generally quite good. Some confusion between single stars and physical binaries is expected because the binary sample includes systems with very large brightness ratios (see section~\ref{sect:msc}). The relatively poor performance on quasars arises mostly due to a confusion with galaxies. If only the photometric subclassifier is used then a similar performance is obtained, but the confusion is then mostly with white dwarfs. Note that these results already assume quasars to be rare (1 for every 100 stars).  Note that for faint, distant stars the true parallax and proper motions can be comparable to the magnitude of the uncertainty in the Gaia measurements, in which case the astrometry does not allow a good discrimination between Galactic and extragalactic objects.

These results should not be over-interpreted, however, as the performance depends strongly on the number of classes included in the training, to the extent that excluding certain classes can lead to much better results. Performance also depends on the relative numbers of sources in each class as well as their parameter distributions in the training and test data sets. Optimizing these is an important part of the on-going work.

\subsection{Stellar parameters from BP/RP (GSP-Phot)}\label{sect:gspphot}

The objective of GSP-Phot is to estimate \teff, \feh, \logg\ and the line-of-sight extinction, \a0, for all single stars observed by Gaia. The extinction is effectively treated as a stellar parameter. (The total-to-selective extinction parameter, \relext, may be added at a later stage.)
GSP-Phot uses the BP/RP spectrum and, in one algorithm, the parallax.
In addition to being part of the Gaia catalogue, the AP estimates are used by several downstream algorithms in Apsis (Figure~\ref{fig:apsisarchitecture}) and elsewhere in the Gaia data processing. 
The algorithms and their performance are described in more detail in 
\citet{2012MNRAS.426.2463L} and the references given below.

GSP-Phot is applied to all sources irrespective of class. While we could exclude those sources which DSC assigns a low star probability, the majority of Gaia sources are stars, so excluding them saves little computing time.  A threshold on this probability can be applied by any user of the catalogue according to how pure or complete they wish their sample to be. 

GSP-Phot contains four different algorithms. Each provides AP estimates for each target:
\begin{enumerate}
\item Priam \citep{LL:DWK-001}: Early in the mission, no calibrated BP/RP spectra are available. Priam uses only the integrated photometry ($G$, \gbp, \grp, \grvs) to estimate \teff\ and \a0\ (see below). This algorithm uses SVM models trained on synthetic spectra.
\item SVM: An SVM is trained to estimate each of the four stellar APs using the BP/RP spectra. SVMs are computationally fast and relatively robust, but we find them not to be the most accurate method for GSP-Phot. Furthermore, a standard SVM does not provide uncertainty estimates (although techniques do exist for extracting these from SVMs).
The SVM AP estimates will also be used to initialize the next two algorithms.
\item {\sc ilium} \citep{2010MNRAS.403...96B}: This uses a forward model, fit using labelled data, to predict APs given the observed BP/RP spectrum.
An iterative Newton--Raphson minimization algorithm is used to find the best fitting forward-modelled spectrum, and thus the APs and their covariances. A two-component forward model is used to retain sensitivity to
the ``weak'' APs \logg\ and \feh\, which only have a weak impact on the stellar spectrum compared to \teff\ and \a0.
\item Aeneas \citep{2011MNRAS.411..435B}: This is a Bayesian method employing a forward model and a Monte Carlo algorithm to sample the posterior probability density function over the APs, from which parameter estimates and associated uncertainties are extracted. Aeneas may be applied to the BP/RP spectrum alone, or together with the parallax. When using the parallax, the algorithm demands (in a probabilistic sense) that the inferred parameters be consistent not just with the spectrum, but also with the parallax and apparent magnitude. Consistency with the Hertzsprung-Russell diagram can also be imposed, thereby introducing constraints from stellar structure and evolution.
\end{enumerate}
As single stars are the main Gaia target, we decided that multiple algorithms and multiple sets of AP estimates were desirable for the sake of consistency checking.  Tests to date show that SVM, {\sc ilium}, and Aeneas are each competitive in some part of AP space or SNR regime. Our plan is that individual results as well as a single set of ``best'' APs for each source will be published in the data releases (see section~\ref{sect:multipleAPs}). 
Nonetheless, we may find during the data processing that some or all algorithms are unable to provide useful estimates of ``weak'' APs on fainter stars.

The APs provided by GSP-Phot are of course tied to the stellar libraries on which it was trained, and different libraries may produce different results. As no single library models the full AP space better than all others, we work with multiple libraries. We could attempt to merge all libraries into one, but this would hide the resulting inhomogeneities (or even introduce errors). We decided instead to train a GSP-Phot model on each library independently, and use each to estimate
APs for a target source. The most appropriate set of results (i.e.\ library) can be decided post hoc based either on a model comparison approach, the estimated uncertainties, or perhaps a simple colour cut. This is still under investigation.

\begin{figure}
\includegraphics[width=8.3cm]{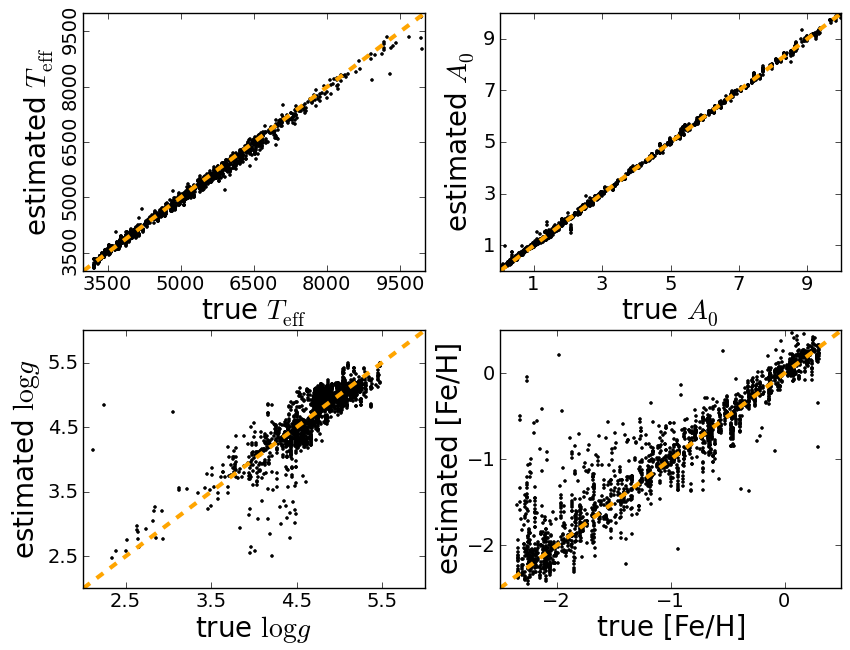}
\caption{Accuracy of AP estimation with the GSP-Phot algorithm Aeneas using
BP/RP spectra of stars at $G$\,=\,15 covering the full AP space shown.
There are a total of 2000 stars in this test sample. The vertical
structure of points visible in the bottom panels is due to our procedure
to generate test spectra from a limited supply of synthetic spectra. This
causes spectra of identical AP values to appear multiple times in the test
set, though these spectra differ in their noise realizations.
\label{fig:GSP-Phot:AP-results-PHOENIX}
}
\end{figure}

Since the publication of GSP-Phot results by \citet{2012MNRAS.426.2463L}, SVM and in particular Aeneas have been improved. Figure~\ref{fig:GSP-Phot:AP-results-PHOENIX} and Table~\ref{table:GSP-Phot:AP-results-PHOENIX} summarize the current internal accuracy of Aeneas, using parallaxes as well as BP/RP.

\begin{table}
\begin{center}
\caption{Accuracy of AP estimation (internal RMS errors) with the GSP-Phot algorithm Aeneas using
BP/RP spectra and parallaxes, for stars covering the full AP space shown in Figure~\ref{fig:GSP-Phot:AP-results-PHOENIX}.
\label{table:GSP-Phot:AP-results-PHOENIX}
}
\begin{tabular}{crrccc}
\toprule
& $G$ & \teff\ & \a0\ & \logg\ & \feh\ \\
& mag & K & mag & dex & dex \\
\midrule
\multirow{3}{*}{\begin{sideways}\begin{footnotesize}A stars\end{footnotesize}\end{sideways}} 
 & 9 & 340 & 0.08 & 0.43 & 0.86 \\
 & 15 & 260 & 0.06 & 0.38 & 0.93 \\
 & 19 & 400 & 0.15 & 0.51 & 0.74 \\
\midrule
\multirow{3}{*}{\begin{sideways}\begin{footnotesize}F stars\end{footnotesize}\end{sideways}}
 & 9 & 150 & 0.06 & 0.36 & 0.36 \\
 & 15 & 170 & 0.07 & 0.38 & 0.33 \\
 & 19 & 630 & 0.35 & 0.37 & 0.60 \\
\midrule
\multirow{3}{*}{\begin{sideways}\begin{footnotesize}G stars\end{footnotesize}\end{sideways}}
 & 9 & 140 & 0.07 & 0.31 & 0.14 \\
 & 15 & 140 & 0.07 & 0.22 & 0.16 \\
 & 19 & 450 & 0.33 & 0.45 & 0.65 \\
\midrule
\multirow{3}{*}{\begin{sideways}\begin{footnotesize}K stars\end{footnotesize}\end{sideways}}
 & 9 & 100 & 0.09 & 0.26 & 0.19 \\
 & 15 & 90 & 0.08 & 0.26 & 0.21 \\
 & 19 & 230 & 0.23 & 0.36 & 0.48 \\
\midrule
\multirow{3}{*}{\begin{sideways}\begin{footnotesize}M stars\end{footnotesize}\end{sideways}}
 & 9 & 60 & 0.13 & 0.15 & 0.21 \\
 & 15 & 70 & 0.14 & 0.29 & 0.25 \\
 & 19 & 90 & 0.13 & 0.17 & 0.29 \\
\bottomrule
\end{tabular}
\end{center}
\end{table}

As noted above, the purpose of Priam is to characterize the stars in the early data releases only, before BP/RP is calibrated.
As the $G-$\gbp\ and $G-$\grp\ colours are almost perfectly correlated, these three bands yield essentially 
just one colour, making it impossible to estimate two APs (\teff\ and \a0) without using prior information. 
Assuming \a0\,$<2$\,mag (but with \teff\,=\,3000--10\,000\,K),
we can estimate \teff\ and \a0\ to an RMS accuracy of 1000\,K and 0.4\,mag respectively using three bands (the latter decreases to 0.3\,mag if we introduce \grvs).
If we can assume \a0\,$<0.1$\,mag, then \teff\ can be estimated to an accuracy of 550\,K using either three or four bands.

Aeneas has also been tested on real data. It was used by \citet{2011MNRAS.411..435B} to estimate \teff\ and \a0\ for 50\,000 Hipparcos FGK stars cross-matched with 2MASS, using the parallax and five band photometry (two from Hipparcos, three from 2MASS). The forward model was fit to a subset of the observed photometry, with temperatures obtained from echelle spectroscopy, and extinction modelled by applying an extinction law to the photometry. \teff\ and \a0\ could be estimated to precisions of 200\,K and 0.2\,mag respectively, from which a new HRD and 3D extinction map of the local neighbourhood could be constructed.

\subsection{Stellar parameters from RVS (GSP-Spec)}\label{sect:gspspec}

GSP-Spec estimates \teff, \logg, global metallicity \mh, alpha element abundance \aabun, and some individual chemical abundances for single stars using continuum-normalized RVS spectra (i.e.\ each spectrum is divided by an estimate of its continuum). Source selection is based on the DSC single star probability, and GSP-Spec can optionally use the measured rotational velocities (\vsini) from CU6, as well as the the stellar parameters from GSP-Phot.

Presently, three algorithms are integrated in the GSP-Spec module: MATISSE \citep{2006MNRAS.370..141R}, DEGAS \citep{2011A&A...535A.106K,2012StatMethodBijaoui}, and GAUGUIN \citep{2012StatMethodBijaoui}.
MATISSE is a local multi-linear regression method. The stellar parameters are determined through the projection of the input spectrum on a set of vectors, calculated during a training phase. The DEGAS method is based on an oblique k-d decision tree. GAUGUIN is a local optimization method implementing a Gauss--Newton algorithm, initialized by parameters determined by GSP-Phot or DEGAS. The algorithms perform differently in different parts of the AP and SNR space. Which results will be provided by which algorithm will be decided once we have experience with the real Gaia data.
As the estimation of the atmospheric parameters and individual abundances from RVS is sensitive to the pseudo-continuum normalization, GSP-spec renormalizes the RVS spectra through an iterative procedure coupled with the stellar parameters as determined by the three algorithms \citep{2011A&A...535A.106K}.

\begin{table}
\begin{center}
\caption{Accuracy of AP estimation (internal RMS errors) with GSP-Spec for RVS spectra for selected AP ranges.
Thin disk dwarfs are defined as \logg\,$>$\,3.9\,dex and $-0.5<$\,[M/H]\,$<-0.25$\,dex, thick disk dwarfs as
\logg\,$>$\,3.9 and $-1.5<$\,[M/H]\,$<-0.5$\,dex, and halo giants as $4000<$\,\teff\,$<6000$\,K, \logg\,$<$\,3.5\,dex and $-2.5<$\,[M/H]\,$<-1.25$\,dex.
\label{tab:gspspec}
}
\begin{tabular}{crrcc}
\toprule
& \grvs\ & \teff\ & \logg\ & \mh\ \\
& mag & K & dex & dex \\
\toprule
\multirow{3}{*}
{\begin{sideways}\begin{footnotesize}Thin\end{footnotesize}\end{sideways}
\begin{sideways}\begin{footnotesize}disk\end{footnotesize}\end{sideways}
\begin{sideways}\begin{footnotesize}dwarfs\end{footnotesize}\end{sideways}} 
 & 10 & 60  & 0.08  & 0.09 \\
 & 13 & 70 & 0.12  & 0.09 \\
 & 15 & 270  & 0.39  & 0.30 \\
\midrule
\multirow{3}{*}
{\begin{sideways}\begin{footnotesize}Thick\end{footnotesize}\end{sideways}
\begin{sideways}\begin{footnotesize}disk\end{footnotesize}\end{sideways}
\begin{sideways}\begin{footnotesize}dwarfs\end{footnotesize}\end{sideways}} 
 & 10 & 70  & 0.11  & 0.09 \\
 & 13 & 110  & 0.17  & 0.12 \\
 & 15 & 350 & 0.43  & 0.29 \\
\midrule
\multirow{3}{*}
{\begin{sideways}\begin{footnotesize}Halo\end{footnotesize}\end{sideways}
\begin{sideways}\begin{footnotesize}giants\end{footnotesize}\end{sideways}}
 & 10 & 70  & 0.17  &  0.15 \\
 & 13 & 90  & 0.28  &  0.17\\
 & 15 & 340  & 0.86  & 0.38 \\
\bottomrule
\end{tabular}
\end{center}
\end{table}

Performance estimates for GSP-Spec are shown in Table~\ref{tab:gspspec}.\footnote{These results are based on a slightly broader RVS pass band extending to 874\,nm. A recent change in the RVS filter has cut this down to 871\,nm. This excludes the Mg lines, which may affect these results and others using RVS quoted in this article.}  
The individual abundances of several elements (Fe, Ca, Ti, Si) will be measured for brighter stars. Based on experience with the Gaia-ESO survey  \citep{2012Msngr.147...25G}, we expect to achieve an internal precision of 0.1\,dex for \grvs\,$<$\,13.

The parameterization algorithms in GSP-Spec have been applied to real data. MATISSE and DEGAS were used in a study of the thick disk outside the solar neighbourhood (700 stars) \citep{2011A&A...535A.107K} and were used in the upcoming final data release (DR4) of the RAVE Galactic Survey (228\,060 spectra). These two applications share almost the same wavelength range and resolution as RVS. MATISSE is used in the AMBRE project \citep{2012A&A...544A.126D} to determine the parameters \teff, \logg, \mh, and \aabun\ of high resolution stellar spectra in the ESO archive (see \citealp{2012A&A...542A..48W} and other forthcoming publications).  MATISSE has also been used to characterize fields observed by CoRoT (\citealp{2013A&A...550A.125G}, \citealp{2010A&A...523A..91G}) and is one of the algorithms being used to characterize FGK stars in the Gaia-ESO survey.

\subsection{Special treatment for emission line stars (ESP-ELS)}\label{sect:espels}

The ELS module classifies emission-line stars, presently into seven discrete classes:
PNe (planetary nebulae), WC (Wolf-Rayet carbon), WN (Wolf-Rayet nitrogen), dMe, Herbig AeBe, Be, and unclassified.
Since some types of emission line star may deserve further treatment in the ESP-HS or ESP-CS 
modules, ESP-ELS is the first of the ESP modules to be applied to the data.
ESP-ELS is triggered by receiving a classification label of ``star'' or ``quasar'' from DSC (the latter included in order to accommodate misclassification in DSC).
The algorithm works on several characteristic features in the BP/RP and/or RVS spectra.  These are centered on (wavelengths in nm): H$\rm{\alpha}$, H$\rm{\beta}$, P14, \ion{He}{ii} $\lambda$468.6, \ion{C}{iii} $\lambda$866, \ion{C}{iv} $\lambda$580.8 and 886, \ion{N}{iv} $\lambda$710, \ion{O}{iii} $\lambda$500.7, and the \cat\ triplet in RVS.  For each of these features a spectroscopic index is defined which minimizes sensitivity to interstellar reddening and instrumental response.  We use the index definition described in \cite{2001MNRAS.326..959C}.  
If significant emission is detected in one or more of the indices, the source is classified using one or
more methods, including a neural network, k-nearest neighbours, and
an interactive graphical analysis of the distribution of various combinations of indices in two-dimensional diagrams.
In this last case, a comparison with the distribution of the indices for template objects is then used to manually define optimal classification boundaries.

The set of template indices was constructed both from synthetic spectra, mainly of
non-emission stars and quasars, and from observed spectra of various types of emission lines stars collected from public telescope archives and online catalogues, 
and supplemented with dedicated ground-based observations. 
The resulting spectral library comprises 1620 spectra of stars belonging to 12 different ELS 
classes (Be, WN, WC, dMe, RS CVn, Symbiotic, T Tauri, Herbig AeBe, Pre-MS, Carbon Mira, 
Novae, PNe) and observed between 320 and 920\,nm \citep{LL:AJL-001}. The spectra were processed with GOG
from which the indices described above were derived.

Typical results of our classification with this template set are shown in Table~\ref{tab:espelsoverview}.
The initial selection thresholds on these indices were set to avoid processing 
non-emission line stars or quasars, with the risk that certain weak emission line stars
will be excluded. This conservative approach, combined with the 
limited resolution and sensitivity drop in the blue wing of the RP H$\rm{\alpha}$
line, leads to not detecting about half of the Be and Herbig AeBe stars. 
Most of the other misclassifications and false detections are due to overlapping spectroscopic 
index values.
Using Gaia observations of a predefined list of known emission line stars, we hope to be able to improve this performance 
and to expand the number of emission line star classes during the mission.

\begin{table}
\begin{center}
\caption{Example of ESP-ELS classification performance in terms of a confusion matrix. 
The rows indicate the true classes, the columns the ESP-ELS assigned class. Each cell gives the percentage of objects classified from each true class to each ESP-ELS class. The dashes indicate exactly zero. Uncl. = Unclassified; Star = Star without emission.
\label{tab:espelsoverview}
}
\footnotesize{
\begin{tabular}{lrrrrrrrr}
\toprule
 True   & \multicolumn{8}{c}{Output class}\\
 class  & PNe  & DMe   & AeBe & Be   & WC   & WN & Uncl.\ & Star \\
 \midrule
PNe       & 63 &  $-$   &  $-$   &  $-$   &  $-$   & $-$ & 28 & 9\\
DMe      &   $-$  & 60 &  $-$   &  $-$   &  $-$   &  $-$  &  $-$  & 40 \\
AeBe      &   $-$  &  $-$   & 48 & 9 &  $-$   &  $-$  &  $-$  & 43 \\
Be        &   $-$  &  $-$   & 5 & 41  &  $-$   &  $-$  &  $-$  & 54 \\
WC        &   $-$  &  $-$   &  $-$   &  $-$    &  74  & 1  & 21  & 4 \\
WN        &   $-$  &  $-$   &  $-$   &  $-$    &  1  & 73  & 18  & 8  \\
\bottomrule
\end{tabular}
}
\end{center}
\end{table}

\subsection{Special treatment for hot stars (ESP-HS)}\label{sect:esphs}

Emission lines in hot (OB) stars will confuse AP estimation methods which assume the entire spectrum has a temperature-based origin in the photosphere. As emission lines are difficult to model reliably, the ESP-HS package attempts to improve the classification of hot stars by omitting those regions of the BP/RP and RVS spectra dominated by emission lines.  The comparison with the template spectra over the selected regions is achieved with a minimum distance method using simplex minimization, while error bars are derived in a second iteration by computing the local covariance matrix. This approach is similar to the one used by \citealp{2006A&A...451.1053F}. The spectral regions to omit are selected based on the results of ESP-ELS. If that module detects emission, then those regions most affected by emission (for that ELS class) are omitted, otherwise the full spectrum is used.

ESP-HS is applied to all stars previously classified by GSP-Phot or GSP-Spec as early-B and O-type stars (specifically \teff\,$>$14\,000\,K). RVS spectra are used for sources which have \grvs$<12$, otherwise only BP/RP spectra are used. This may be extended to fainter magnitudes during the mission depending on the quality of the RVS spectra.
ESP-HS always estimates \teff, \logg, and \a0. Assuming the BP spectrum is available \feh\ is also estimated.
If an RVS spectrum is available and \vsini\ has not been estimated already by CU6, ESP-HS will derive this too.

We applied the algorithm on spectra randomly spread over the parameter space (\teff\,$>$14\,000\,K).  The maximum fractional residuals we found are given in Table~\ref{tab:esphsoverview}. The current version of the algorithm is unable to correctly derive the APs from BP/RP for stars fainter than $G=18$. During the mission, the results will be validated by comparing our derived APs to those obtained for a sample of reference O, B and A-type stars \citet{LL:AJL-002}. Spectra for these are being collected and analysed as part of both the HERMES/MERCATOR project \citep{2011A&A...526A..69R} and the Gaia--ESO Survey.

\begin{table}
\begin{center}
\caption{Maximum fractional AP residuals, i.e.\ |measured-true|/true, for the ESP-HS algorithm as a function
of the G magnitude. $N$ is the number of cases for each magnitude range.\label{tab:esphsoverview}}
\footnotesize{
\begin{tabular}{r@{}c@{}lrrrr}
\toprule
\multicolumn{3}{c}{{\sc G}} & $\Delta$\teff\ & $\Delta$\logg\ & $\Delta$\a0\ & $N$ \\
\midrule
0&--&10 & 0.10 &  0.15 &  0.08 & 504 \\
10&--&15 & 0.17 &  0.31 &  0.12 & 1154 \\
15&--&18 & 0.25 &  0.40 &  0.55 & 1102 \\
\bottomrule
\end{tabular}
}
\end{center}
\end{table}

\subsection{Special treatment for cool stars (ESP-CS)}\label{sect:espcs}

\begin{figure*}
\begin{center}
\includegraphics[width=0.8\textwidth]{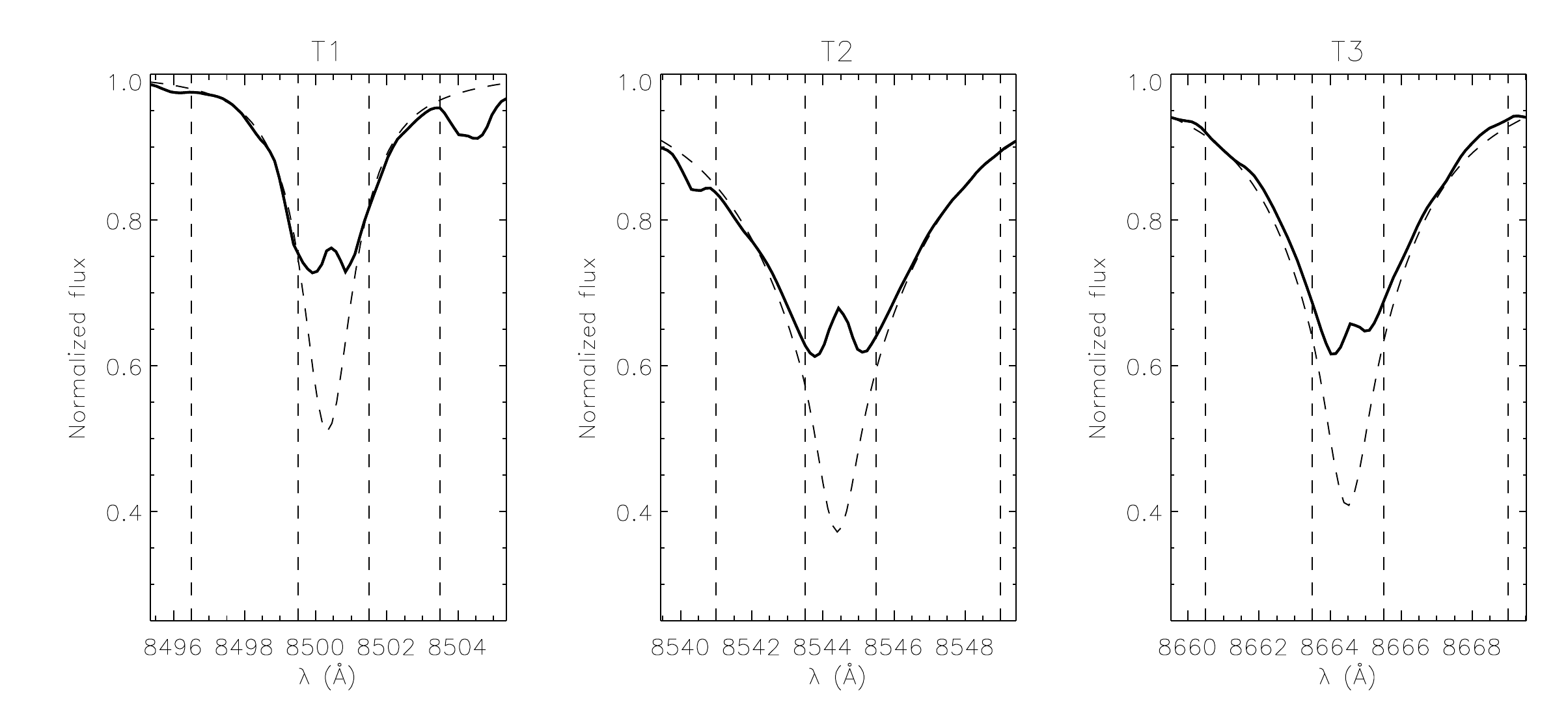}
\caption[]{The three parts of the \cat\ triplet of the active star HD82443 (solid line) 
observed at high resolution by TNG/SARG and processed by GOG to simulate an RVS spectrum at the noise level of a $G=8$ source. The dashed curve shows an NLTE synthetic spectrum for a pure photospheric contribution for comparison. Dashed vertical lines delimit the wings and the core regions.}
\label{fig:esp-cs-sample-HD82443}
\end{center}
\end{figure*}

ESP-CS applies procedures for analysing peculiarities arising from magnetic activity and the presence of circumstellar material in stars with \teff\ in the range 2500--7500\,K.

Chromospheric activity can be detected via a fill-in of the \cat\ infrared triplet lines in the RVS spectrum relative to the spectrum of an inactive star (Figure~\ref{fig:esp-cs-sample-HD82443}). Strong emission in the core of these lines for young stars generally indicates mass accretion. The degree of activity can be quantified by subtracting the spectrum of an inactive star from that of an active star with the same parameters, then measuring the difference in the central depth of a line, $R_{\rm IRT}$, or in its integrated absorption, $\Delta W$ \citep{Busa_etal:2007}. (These are the outputs from ESP-CS.) This is challenging because it requires knowledge of the star's radial velocity, rotational velocity (\vsini), \teff, \logg, and \feh. NLTE effects must also be modelled \citep{Andretta_etal:2005}.

ESP-CS nominally adopts APs from GSP-Phot (and possibly GSP-Spec). However, these could be adversely affected by the star's UV- or IR-excess from magnetic activity and/or circumstellar material (and the GSP-Spec estimates could be distorted by high \vsini\ in young active stars). 
For these reasons ESP-CS also estimates stellar APs itself, using $\chi^2$ minimization against a set of templates of the \cat\ wings. This takes into account rotational broadening and is unaffected by photometric excesses, so should provide a higher degree of internal consistency for the activity estimation. This approach is motivated by various studies in the literature showing that the wings of these lines are sensitive to all three stellar APs in some parts of the parameter space \citep{Chmielewski:2000, Andretta_etal:2005}. For metal rich dwarfs, the estimated activity level is actually not very sensitive to these stellar APs. Giants, in contrast, demand a higher accuracy of the APs, to better than 10\% to get even a coarse estimate of chromospheric activity.

Based on existing $R'_{\rm HK}$ catalogues (e.g.\ \citealp{Henry_etal:1996,Wright_etal:2004}), the Besancon galaxy model \citep{Robin_etal:2003}, and the $R_{\rm IRT}$ and $\Delta W$ correlations with $R'_{\rm HK}$ from \cite{Busa_etal:2007}, we predict that Gaia will measure chromospheric activity to an accuracy of 10\% in about 5000 main sequence field stars using $\Delta W$, and in about 10\,000 stars using $R_{\rm IRT}$, which is some five times as many as existing activity measurements. We also expect to be able to measure activity in giants below the Linsky--Haish dividing line \citep{Linsky_Haish:1979}, with the survey's homogeneity being an added value for statistical studies.  Finally, we expect to be able to identify very young low-mass stars in the field and in clusters down to \grvs\,=\,14 via their accretion or chromospheric activity signatures in the Calcium triplet.

\subsection{Special treatment for ultra-cool stars (ESP-UCD)}\label{sect:espucd}

The ESP-UCD module provides physical parameters for the coolest stars observed by Gaia, \teff\,$<$\,2500\,K, hereafter referred to as ultra-cool dwarfs (UCDs) for brevity. The design of the module and its results are described in detail in \cite{2013A&A...550A..44S}, so we limit our description to the main features.

Based on empirical estimates of the local density of ultra-cool field dwarfs \citep{2008A&A...488..181C}, the BT-Settl family of synthetic models \citep{2012RSPTA.370.2765A}, and the Gaia instrument capabilities outlined in section~\ref{sect:gaiadata}, the expected number of Gaia detections per spectral type bin ranges from a few million at M5, down to a few thousand at L0, and several tens at L5. According to the Gaia pre-launch specifications, it should be possible to detect UCDs between L5 and late-T spectral types, although only 10--20 such sources are expected.

The ESP-UCD module comprises two stages: the select and process submodules. The select submodule identifies good UCD candidates for subsequent analysis by the process submodule. This selection is done according to pre-defined and non-conservative cuts in the proper motion, parallax, $G$ magnitude, and \gbp$-$\grp\ colour.  The exact definition of the selection thresholds is based on the BT-Settl grid of synthetic models and is likely to change as a result of the internal validation of the module during the mission. In order to be complete at the hot boundary (2500\,K) of the UCD domain, the module also selects stars which are hotter than this limit but fainter than the brightest UCD (since according to the models, a 2500\,K low gravity star can be significantly brighter than hotter stars with higher gravities). Therefore, the ESP-UCD selection aims to be complete for sources brighter than $G=20$ and up to 2500\,K, but will also contain sources between 2500 and approximately 2900\,K. The module is actually trained on objects with \teff\ up to 4000\,K. During the data processing we may refine the selection and what we define as the hot boundary for the UCD definition.

The process submodule estimates \teff\ and \logg\ from the RP spectrophotometry using three methods for each source: $k$-nearest neighbours \citep{Cover:Hart:1967}, Gaussian Processes \citep{rasmussen2006gaussian}, and Bayesian inference \citep{sivia2006data}.  In all three cases the regression models are based on the relationship between physical parameters and observables defined by the BT-Settl library of models. ESP-UCD will provide all three estimates.  The root-mean-square (RMS) \teff\ error, estimated using spectra of real UCDs obtained with ground-based telescopes and simulated to look like RP spectra at $G=20$, is 210\,K. The lack of estimates of \logg\ for this set of ground-based spectra of UCDs prevents us from estimating the RMS error for this parameter, but the internal cross-validation experiments show an RMS error of 0.5\,dex in \logg, which is probably an underestimate of the real uncertainty.

\subsection{Multiple Star Classifier (MSC)}\label{sect:msc}

\begin{figure}
\begin{center}
\includegraphics[width=0.34\textwidth]{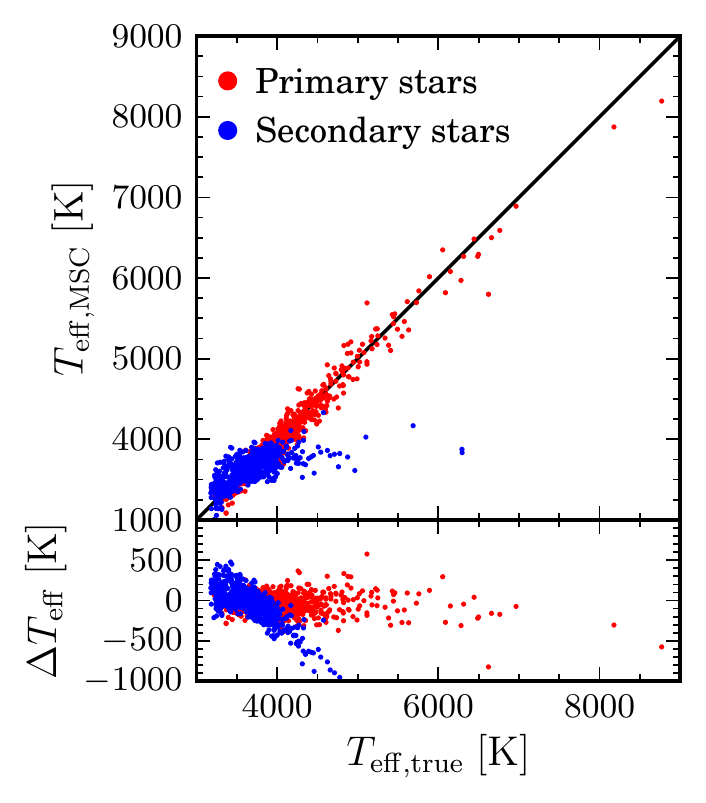}
\caption[]{
The performance of MSC at estimating \teff\ for both components of
an unresolved dwarf binary systems at $G=15$. The upper panel shows the predicted vs.\ true \teff, the lower the residual (predicted minus true) vs.\ true \teff, for the primary component (red) and secondary (blue).
\label{fig:msc_teff}}
\end{center}
\end{figure}

\begin{figure}
\begin{center}
\includegraphics[width=0.34\textwidth]{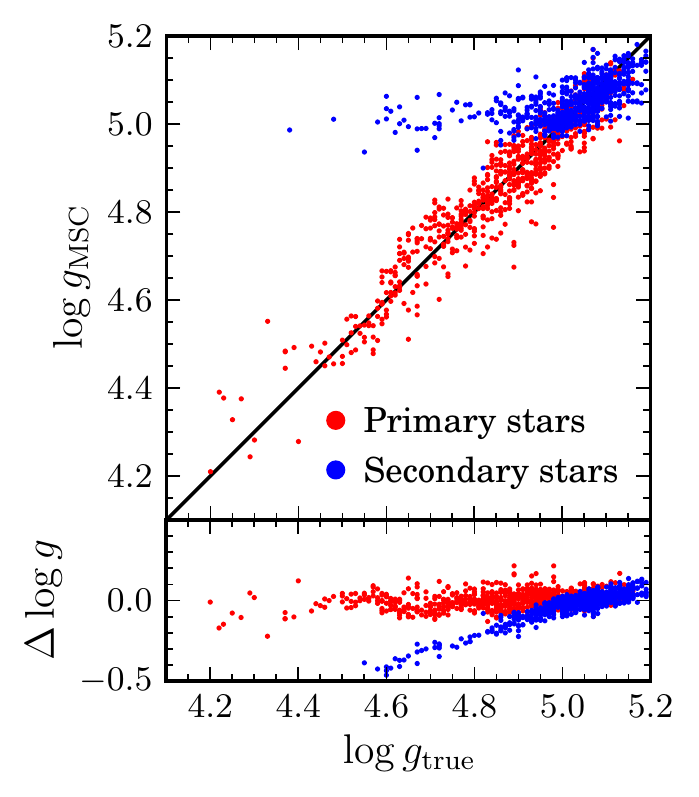}
\caption[]{
As Figure~\ref{fig:msc_teff}, but for the surface gravity.
\label{fig:msc_logg}}
\end{center}
\end{figure}

MSC uses the BP/RP spectrum to estimate the APs of sources identified as unresolved physical binaries by DSC.  Currently it estimates \a0, \feh, and the brightness ratio BR\,=\,$\log_{\rm 10}(L_1/L_2)$, of the system as a whole ($L_1$ and $L_2$ are the component luminosities), as well as the effective temperature and surface gravities of the two components individually.  The APs are estimated using SVMs, using the same SVM implementation used for GSP-Phot (section~\ref{sect:gspphot}). 
For more details see \citet{LL:PAT-009} and \cite{TBJ2012}.

The training and testing data for MSC are constructed by combining spectra for single stars into physically plausible binaries (Lanzafame 2007, private communication), and simulating them with GOG \citep{LL:RS-010}.
A system age and metallicity is selected at random and masses are drawn from a Kroupa IMF
and paired randomly. The corresponding atmospheric parameters
are identified using Padova isochrone models.
We then use the MARCS spectra (section~\ref{sect:training}) to represent the individual stars with the closest corresponding parameters.

\begin{table}
\begin{center}
\caption{Summary of MSC performance in terms of the RMS, median absolute
  residuals (MAR), and median residual (MR, as a measure of
  systematic errors). The subscripts 1 and 2 denote the primary and the secondary components respectively. The apparently good performance for the secondary component is mostly an artefact of the narrow distribution of the APs in our data set.
\label{tab:msc1}}
\begin{tabular}{lrrr}
\toprule
AP & RMS & MAR & MR \\
\midrule
$T_{{\rm eff},1}$ & 130 &  90 &  70 \\
$T_{{\rm eff},2}$ & 260 & 160 & 110 \\
$\log g_1$ & 0.05 & 0.04 & 0.03 \\
$\log g_2$ & 0.10 & 0.06 & 0.03 \\
\feh\ & 0.14 & 0.10 & 0.07 \\
\a0\ & 0.11 & 0.09 & 0.07 \\
$BR$ & 0.36 & 0.26 & 0.19 \\
\bottomrule
\end{tabular}
\end{center}
\end{table}

The performance of MSC depends not only on the magnitude of the system, but also the brightness ratio of the two components. MSC has been trained on a data set of dwarf--dwarf binaries at $G=15$ with an exponentially decreasing distribution in BR from 0 to 5, such that the majority have BR\,$<2$. Yet even at B\,=\,2 the secondary is a hundred times fainter than the primary, so we should not expect good average performance on the secondary component.

Here we report results on a test data set limited to BR$<1.5$. Figure~\ref{fig:msc_teff} shows the \teff\ residuals for both components as a function of \teff.  We see that we can predict \teff\ for the primary star quite accurately, but not for the secondary star, despite the apparently good summary statistics given in Table~\ref{tab:msc1}.
Figure~\ref{fig:msc_teff} shows that for the secondary star, the estimated \teff\ correlates poorly with the true \teff. This is because the SVM is hardly able to learn the generally weak signature of $T_{{\rm eff},2}$ from the data, so assigns
essentially random values from the training data distribution.
The RMS is then relatively small simply because the spread in the $T_{{\rm eff},2}$ in the training data is also small.\footnote{This serves as a reminder that performance as measured by RMS residuals should be judged in comparison to the standard deviation in the data.} 
This narrow $T_{{\rm eff},2}$ distribution is a consequence of the way the binary systems were constructed. 

We see a similar problem with the determination of the surface gravity of the secondary in Figure~\ref{fig:msc_logg}. This is not 
at all surprising, because this is anyway a weak parameter. 
In contrast, \logg\ for the primary can be determined quite accurately (no significant systematics)
despite the interfering spectrum of the secondary. This is partly a consequence of a correlation between \teff\ and \logg\ in our data set. On the other hand, when such correlations are real, they should be exploited.

If we extend the test data set to include systems with larger BR, then the accuracy of both components degrades. 
Conversely, limiting it to smaller BRs produces better average performance.
Although the performance on the secondary components is relatively poor, MSC is nonetheless a useful algorithm because it gives a better performance on the primary components than does GSP-Phot \citep{LL:PAT-009}. That is, neglecting the existence of the secondary degrades the accuracy with which we can estimate the APs of the primary.

MSC will report statistical uncertainties on the predicted APs obtained from the residuals on a test data set: Given the predictions of the APs of an unknown object, we use a look-up table to find the typical errors obtained on a test set around those measured APs.

\subsection{Stellar mass and age estimation (FLAME)}\label{sect:flame}

Using the atmospheric AP estimates from GSP-Phot, it is possible to infer, to a great or lesser accuracy, the fundamental stellar parameters of age and mass. This is the task of FLAME.  First, the luminosity of each single star will be computed from the Gaia parallaxes and magnitudes, and suitable bolometric corrections calculated from the \teff.  The luminosity, \teff\ and \feh\ estimates place the object in the Hertzsprung-Russell diagram.  Comparison of this position with stellar evolutionary tracks for a range of masses and abundances then allows us to estimate the mass, age and metallicity fraction, $Z$, of the star through an inversion method.  The initial helium abundance in mass fraction $Y$ is assumed to follow the helium to metal enrichment law $\frac{\Delta Y}{\Delta Z}= 1.45$, based on the solar model calibration and a primordial helium abundance of $Y = 0.245$. Two different algorithms are used for the inversion (results will be provided from both): a classical $\chi^2$ minimization algorithm \citep{1998A&A...329..943N}, and a forward modelling Bayesian method \citep{2004MNRAS.351..487P} using prior information on the initial mass function (only for age determination), stellar formation rate and metallicity distribution function.  From the results the stellar radius can be calculated. This is important for correcting the zero point radial velocity of stars for the gravitational redshift.

When available, AP estimates (and additional abundance information) from GSP-Spec or ESP could be used by FLAME also.

For stars of A type and later, a 1\% error in \teff\ translates to a similar error in the mass determination. Age is more sensitive due in part to the degeneracy of evolutionary tracks in the HRD, so will have errors of at least 10\%, or even 100\% in the worst case.  This all assumes fixed chemical composition, so the uncertainties will increase when the \feh\ uncertainty is taken into account.

\subsection{Galaxy classification (UGC)}\label{sect:ugc}

We expect to observe a few million unresolved galaxies with Gaia. UGC will use the BP/RP spectra to classify them into discrete classes and to estimate the redshift, the Galactic extinction, and parameters which determine the star formation law in the source galaxy.  We use SVMs for both classification and parameter estimation, the former giving probabilities for each of the galaxy classes. The SVMs are trained on simulated BP/RP spectra generated from the Galaxy spectral libraries described in section~\ref{sect:galspeclib}, which also defines the four galaxy types.

UGC comprises two separate modules \citep{2012BlgAJ..18b...3B}. The first, UGC-Learn, provides SVM tuning, training and testing functions for offline preparation of the models.  A number of SVM models have been trained, arranged in a two-layer hierarchy.  For each of three G magnitude ranges (13--16.5, 16.5--19, 19--20), an SVM in the first layer is trained to cover the total range of the extinction and redshift parameters.  In the second layer, there is a set of SVM models, each dedicated to a narrower range of these two parameters (again for each magnitude range).

The second module, UGC-Apply, applies the fitted SVM models in a hierarchical manner, in two steps. It operates on sources identified by DSC as having a galaxy probability above a predefined threshold (section~\ref{sect:dsc}). In the first step, the ``total range'' SVM for the appropriate source magnitude is applied to provide an initial estimate of the redshift and extinction. In the second step, ``specific range'' SVM models corresponding to these initial parameter estimates are used to classify the spectrum and to estimate the parameters.  The star formation parameters for all the galaxy types are estimated, independently of the best class predicted.

\begin{table}
\begin{center}
\caption{Example of UGC AP estimation performance in terms of the RMS residual for sources at three different magnitudes.
The results from the final step of a two-step approach are given.
\label{tab:ugcExtParams}
}
\vspace*{1em}
\begin{tabular}{lrrr}
\toprule
                    &  \multicolumn{3}{c}{RMS at $G$=}\\
AP                 & 15& 18.5& 20\\
\midrule
\a0\,/\,mag   &  0.04    & 0.15    & 0.35     \\
redshift         & 0.002    & 0.011 & 0.028    \\
\bottomrule
\end{tabular}
\end{center}
\end{table}

\begin{table}
\begin{center}
\caption{Example of UGC classification performance in terms the true positive 
classification percentage. Results for sources of fixed magnitude as well as
for a range of magnitudes are shown. 
\label{tab:ugcGalType}
}
\vspace*{1em}
\setlength{\tabcolsep}{2pt}
\begin{tabular}{lcccccc}
\toprule
{Galaxy}& \multicolumn{6}{c}{True positive percentage at $G$=}                    \\
{type}    & 15    & 13--16.5    & 18.5 & 16.5--19 & 19--20     & 20 \\
\midrule
early    & 93    & 91        & 78    & 79        & 50    & 44 \\
spiral    & 98    & 95         & 90    & 92      & 73    & 64 \\
irregular    & 89    & 86         & 51    & 54      & 26    & 28 \\
quenched~~    & 99    & 98        & 94    & 94      & 83    & 83        \\
\bottomrule
\end{tabular}
\end{center}
\end{table}

UGC shows good performance in predicting the extinction (range 0--6\,mag) and redshift (range 0.0--0.2), as can be seen in Table~\ref{tab:ugcExtParams}. The residuals show no trend with the parameters.  The classification performance is measured as the percentage of true-positive classifications, and is shown in Table~\ref{tab:ugcGalType}.  At all magnitudes the best performance is obtained on the spirals and quenched star formation galaxies. At $G=20$ the true-positive rate for the early and irregular classes is below 50\%, and we are not able to estimate the star formation parameters. For the two brighter magnitude ranges the best performance is achieved for the early type parameters. For irregular and quenched star formation galaxy types the gas infall rate cannot be accurately predicted, even at $G=15$.  Applying the hierarchical approach to these parameters may help.  Performance could be further increased by improving the libraries, representing more realistically the observed spectral types, eliminating overlaps among galaxy types, and providing larger numbers of spectra for the different training and testing data sets. We also plan to investigate using the total Galactic extinction estimates from TGE (section~\ref{sect:tge}).

\subsection{Quasar classification (QSOC)}\label{sect:qsoc}

QSOC processes the 500\,000 or so quasars which we expect Gaia to observe. It has two goals.  First, it classifies each quasar into the three classes type I, type II and BAL (broad absorption line) quasars. This is achieved using a standard SVM classifier. Second, QSOC estimates the redshift, total emission line equivalent width, and the slope of the power law continuum. These three APs are estimated using an ensemble of trees based on the Extremely Randomized Tree (ERT) algorithm \citep{Geurts:2006:ERT:1132034.1132040}.  The redshift is also estimated using an SVM in classification mode, in which each redshift bin of width 0.1 in redshift from z=0 to 5.6 is represented as a separate class and assigned a probability.  (It is done this way to search for possible multimodality in redshift estimation.)  The SVM and ERT redshift estimates are then combined to give a single redshift estimate. The models are trained using the quasar spectral libraries described in section~\ref{sect:training}.

The module has been evaluated using K-fold cross validation. The results reported here are for training and testing with the semi-empirical library.  The SVM classifier achieves an accuracy of 97.4\%, 95.8\%, and 91\% for $G=15$, 18.5, and 20 (respectively), for those quasars where the SVM redshift classifier gives a probability above some threshold in a single bin (i.e.\ high confidence).  The AP predictions from each regression tree in the ERT ensemble could be combined in several different ways. While the mean minimizes the RMS, the median minimizes the mean absolute deviation and reduces the bias.  We instead calculate a discretized mode: we form a histogram of the estimates (with an adapted bin width), and report the central value of the highest bin as our AP estimate.  We find that this gives a higher accuracy than either the mean or the median.  The redshift accuracy obtained in this way is $0.02$, $0.03$, and $0.04$ for $G=15$, 18.5, and 20 respectively. 
We observe some systematically wrong redshift estimates due to degeneracies in the identification of the rest wavelengths of pairs of strong emission lines, as well as some minor biases at small redshifts caused by the zero limit of the redshifts.

So far we do not try to deal with misclassifications from DSC, e.g.\ contamination from Be or WR stars (although by using parallaxes and proper motions, DSC attempts to minimize such misclassifications). We also do not take into account interstellar extinction (the training data, from SDSS DR9, are at high Galactic latitude and assumed to be extinction free), so QSOC may not produce reliable results at low latitudes.  Part of the on-going development will be to use extinction estimates from TGE to overcome this limitation.

\subsection{Total Galactic Extinction (TGE)}\label{sect:tge}

When estimating the APs of objects from their broad band spectra, it is important to take into account the impact of interstellar extinction. Some algorithms, such as GSP-Phot, estimate this for each star independently, effectively treating the extinction as an additional stellar parameter. 
But this approach works less well for some types of object, in particular quasars.  The role of TGE is to estimate the total Galactic extinction (the extinction integrated to the edge of the Galaxy) towards an extragalactic source. It does this by combining the individual extinction estimates from GSP-Phot for distant stars (small parallaxes) in the source's direction.

As part of the DPAC processing, Gaia sources will be indexed according to their position on the sky using the 
Hierarchical Equal Area isoLatitude Pixelisation (HEALPix)
scheme \citep{2005ApJ...622..759G}. This hierarchically partitions the celestial sphere into ever smaller levels, or HEALpixels.  Apsis processes sources in blocks of data corresponding to level 6 HEALpixels, each of which covers 0.839 square degrees.  Given all the data in one such HEALPix, TGE first selects stars to use as extinction tracers. It then uses the estimated extinctions and parallaxes of these tracers to derive the total Galactic extinction for that HEALPix. This is repeated for all HEALpixels over the whole sky.  In regions where there are more tracers, a higher HEALPix level can be used in order to achieve a higher angular resolution map.

Candidate tracers are selected to be single stars (using the DSC probability) and non-variable (using indices from the photometric processing in CU5). Those with the most precise \a0\ estimates are selected based on the expected performance of GSP-Phot in different parts of the AP space and on the individual \a0\ uncertainty estimates provided by GSP-Phot. Finally, a parallax selection criterion is applied to the candidate tracers in order to select just those which are sufficiently far from the main gas and dust layer in the Galactic plane.

For the estimate of the total Galactic extinction for that HEALpixel, \atge, we report the mean \a0\ value of the selected tracers.  The uncertainty in this we represent with the RMS of the \a0\ values.  We are exploring the use of other estimators which enable more robust estimates of \atge, in particular at low Galactic latitudes.  On account of the very high extinction in some fields at low latitudes, there will be insufficient distant tracers for TGE to make a reliable estimate of \atge\ (although we also then expect to detect fewer extragalactic objects on account of this same extinction).

\begin{figure}
\begin{center}
\includegraphics[width=0.45\textwidth]{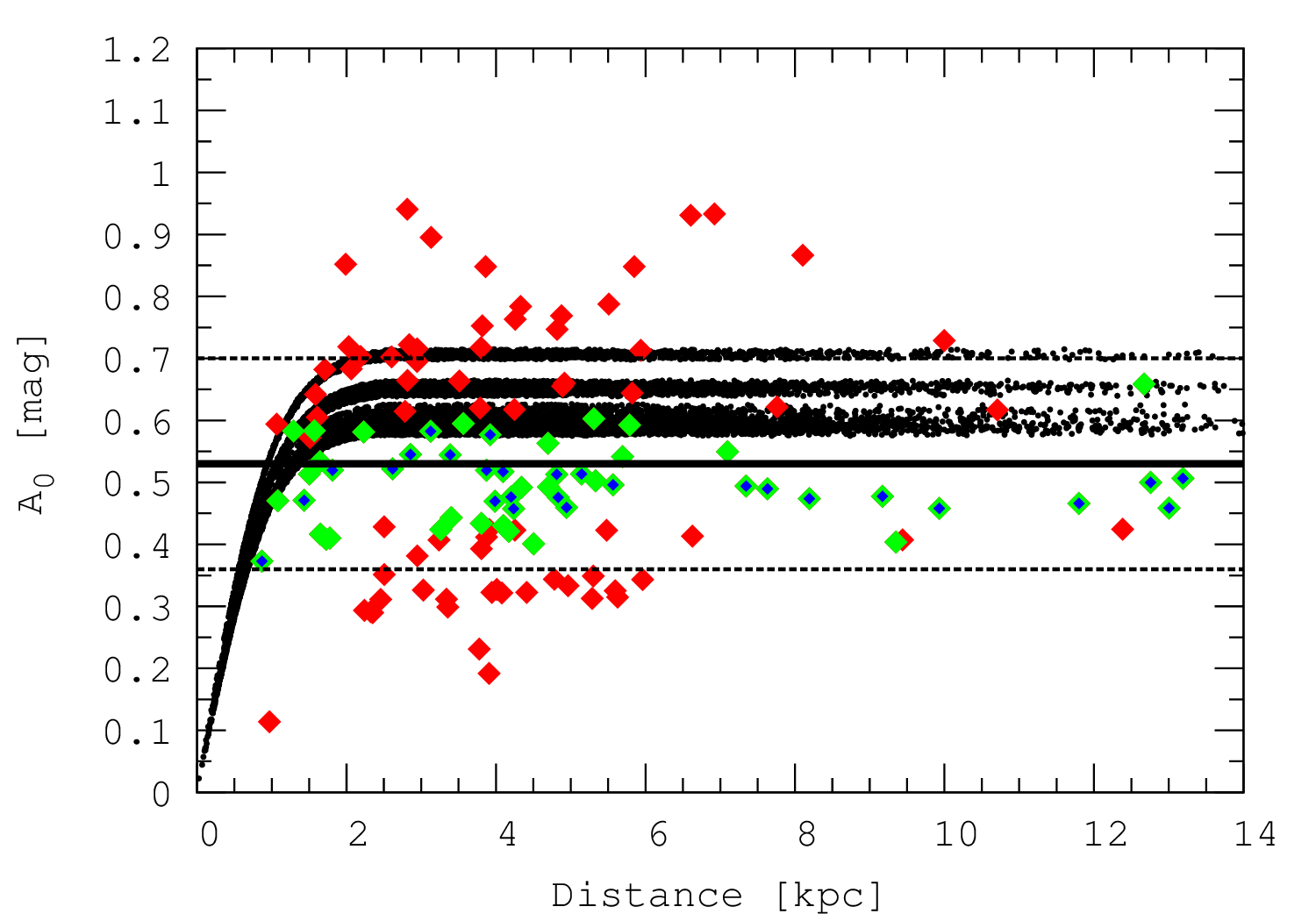}
\caption[]{
Example of TGE extinction estimate in a simulated Galactic field at $(l,b)$\,=\,60\deg,\,10\deg. The black points show the
true extinction, \a0, for all stars in the field, while the diamonds show the estimated extinction from
GSP-Phot for the selected tracers, coloured according to the magnitude of the uncertainty of the \a0\ estimates:
red for $|\Delta A_{\rm 0}| > 0.1$, green for $|\Delta A_{\rm 0}| < 0.1$, and blue for $|\Delta A_{\rm 0}| < 0.05$ mag.
The solid horizontal line shows the estimated  \atge\ value and the dashed lines the $\pm 1\sigma$ RMS.
\label{fig:tgefigure1}
}
\end{center}
\end{figure}

As an example of how TGE works, Figure~\ref{fig:tgefigure1} shows the estimated \a0\ for stars and selected tracers in a particular HEALpixel (from Aeneas in GSP-Phot), using simulated Galactic data, and the value of \atge\ estimated from these. In addition to TGE providing extinction estimates for use in QSOC (and possibly UGC), it will also provide a two-dimensional map of the total extinction for most of the Galaxy, unique in that it will be derived from the individual extinction estimates of stars with measured parallax.

\subsection{Outlier Analysis (OA)}\label{sect:oa}

Like any supervised classification algorithm, DSC can only reliably classify objects which are modelled accurately in its training set. By design, objects which do not match its training set achieve low probabilities for all classes and will therefore be labelled as unknowns, or outliers. These could be types of objects omitted from the DSC training data entirely, objects with poor spectral models, or instrumental artefacts not well modelled by GOG. They could also be previously unseen types of object.  We estimate that 5\% or more of the Gaia sources will be marked as outliers by DSC -- more than 50 million objects -- so some kind of automated analysis of these is mandatory. This is the task of OA. Its main purpose is to help improve the source and instrument modelling and thereby improve the training data sets during the mission.

OA uses a Self-Organizing Map (SOM, \citealp{SelfOrganizingMaps}). This projects the original data (BP/RP and astrometry) into a 2D grid of nodes in a way that attempts to preserve local topology, thereby clustering together similar objects which may be systematically rejected by DSC \citep{FustesSOM2012}.  Then follows an identification stage, where we try to discover whether any other known types of source are associated with any of these clusters. This stage could make use of data from other surveys and catalogues.
 
\begin{figure}
\centering
\includegraphics[width=0.30\textwidth]{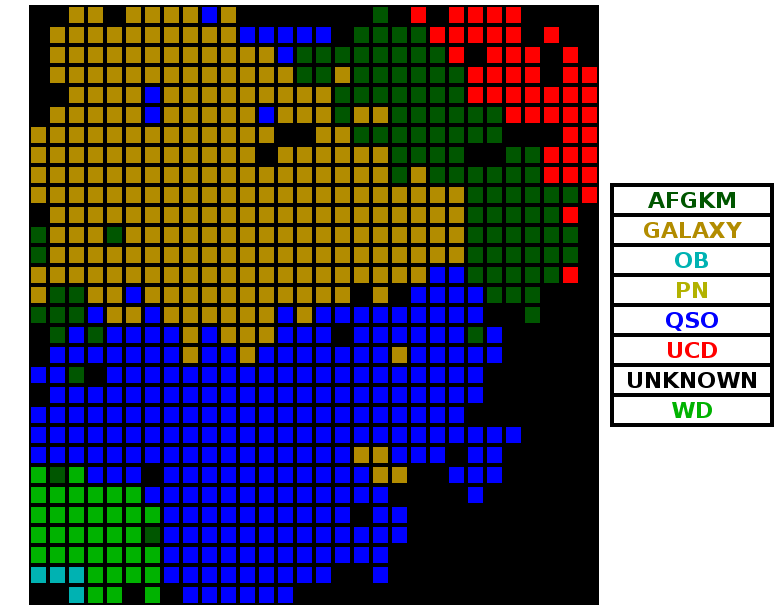} 
\includegraphics[width=0.30\textwidth]{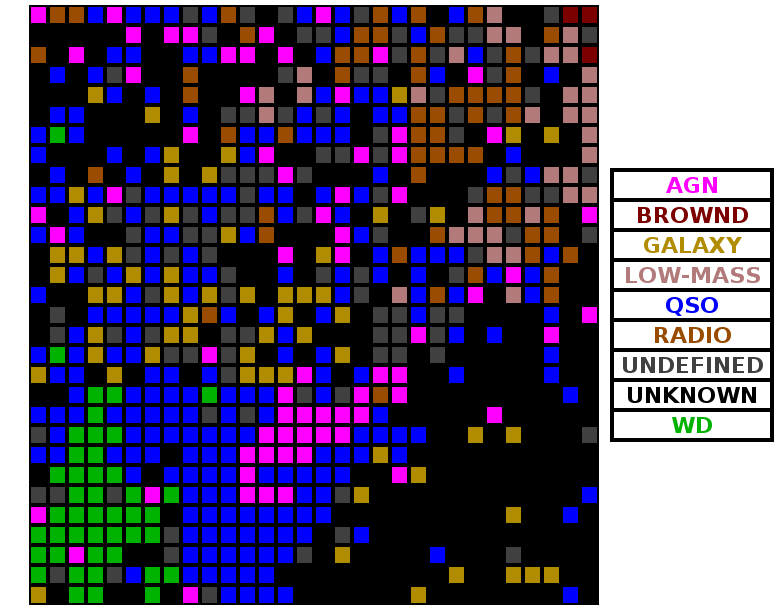} 
\caption{Identifications of SDSS outliers using the SOM in the OA module, obtained from Gaia simulations (top) and Simbad (bottom).
\label{fig:sommaps}
}
\end{figure}

In order to study the behaviour of our algorithm with a realistic dataset, we compiled a semi-empirical BP/RP library from spectra that were classified as ``unknown'' by the SDSS spectroscopic classification pipeline. This dataset comprises 10\,125 objects, which are mostly faint objects, incomplete spectra, or the result of a poor fibre alignment. We fit a SOM with 30$\times$30 nodes to these. As a first method of identification, we applied a k-nearest neighbours classifier on labelled objects from the simulated Gaia data (described in section~\ref{sect:training}). Observing where these tend to land in the SOM, we used this to label the SOM nodes. The results of this are shown in the upper panel of Figure~\ref{fig:sommaps}. (Note that there is no physical meaning to the axes, size or shape of the SOM.) In a second method, we identified from the Simbad database the nearest object on the sky to each SDSS object and retrieved its Simbad class, if available (which is the case for about 3000 of the SDSS objects). If a large enough fraction of objects in a single SOM node share the same class, we label that node with that class, as shown in the lower panel of Figure~\ref{fig:sommaps}. Otherwise the node is labelled ``unknown''. Comparing the two maps, we see that quasars and white dwarfs are identified in similar regions in both cases, giving some confidence that these classifications are appropriate. Using this approach, we were able to identify 400 white dwarf candidates, 1000 quasar candidates, and 16 brown dwarf candidates from among the SDSS outliers. For further details, see \cite{AA/2013/21445}.
 
\subsection{Unsupervised clustering (OCA)}\label{sect:oca}

The supervised learning modules used in Apsis can only assign meaningful parameters to the types of objects they have been trained on. Inevitably, not all types of objects which Gaia will encounter are covered by the Apsis training sets. 
The aim of the Object Clustering Algorithm (OCA) is to identify some natural groups among the Gaia sources, independently of labelled training data sets. This can be used to help improve the training sets in a similar manner to OA, but may also identify potentially new types of object which can then be studied further.\footnote{We use the standard term ``natural groups'' here, but in fact there is no such thing. The clustering found by any algorithm is determined by the similarity measure -- or distance metric -- we choose to adopt.}

OCA implements a variant of the Hierarchical Mode Association Clustering  algorithm (HMAC, \citealp{jmlr:LRL-2007}), which analyses the density of sources in the multidimensional space formed by the data (here BP/RP, astrometry and, when available, RVS). In this framework, each individual source is associated to the closest mode (maximum) of the probability density landscape. Rather than explicitly computing the probability density, the Modal Expectation-Maximisation (MEM) algorithm is used to assign sources to modes.  This works by climbing to a local maximum in an iterative fashion. It is almost equivalent to assigning the source to the closest mode in a kernel density estimation \citep{book:WJ}.  The algorithm uses Gaussian kernels to find the modes, and it becomes hierarchical when we increase the size of the covariance matrix of these kernels, such that modes (and their associated sources) are merged into new modes/clusters at higher levels of the hierarchy.

The computational complexity of HMAC is quadratic in the number of sources, which is up to $10^9$ in the case of Gaia. In order to fit within computational time and memory limitations, we use a divide-and-conquer strategy. The full sample of sources to be clustered is partitioned into disjoint subsamples corresponding to different HEALpixels on the celestial sphere. HMAC is applied to each of these subsamples in order to identify modes (cluster representatives). These modes are assigned a weight proportional to the number of sources that converged to it (the cluster size). An iterative process is then used to merge the modes across subsamples.

\begin{figure}
\begin{center}
\includegraphics[width=0.4\textwidth]{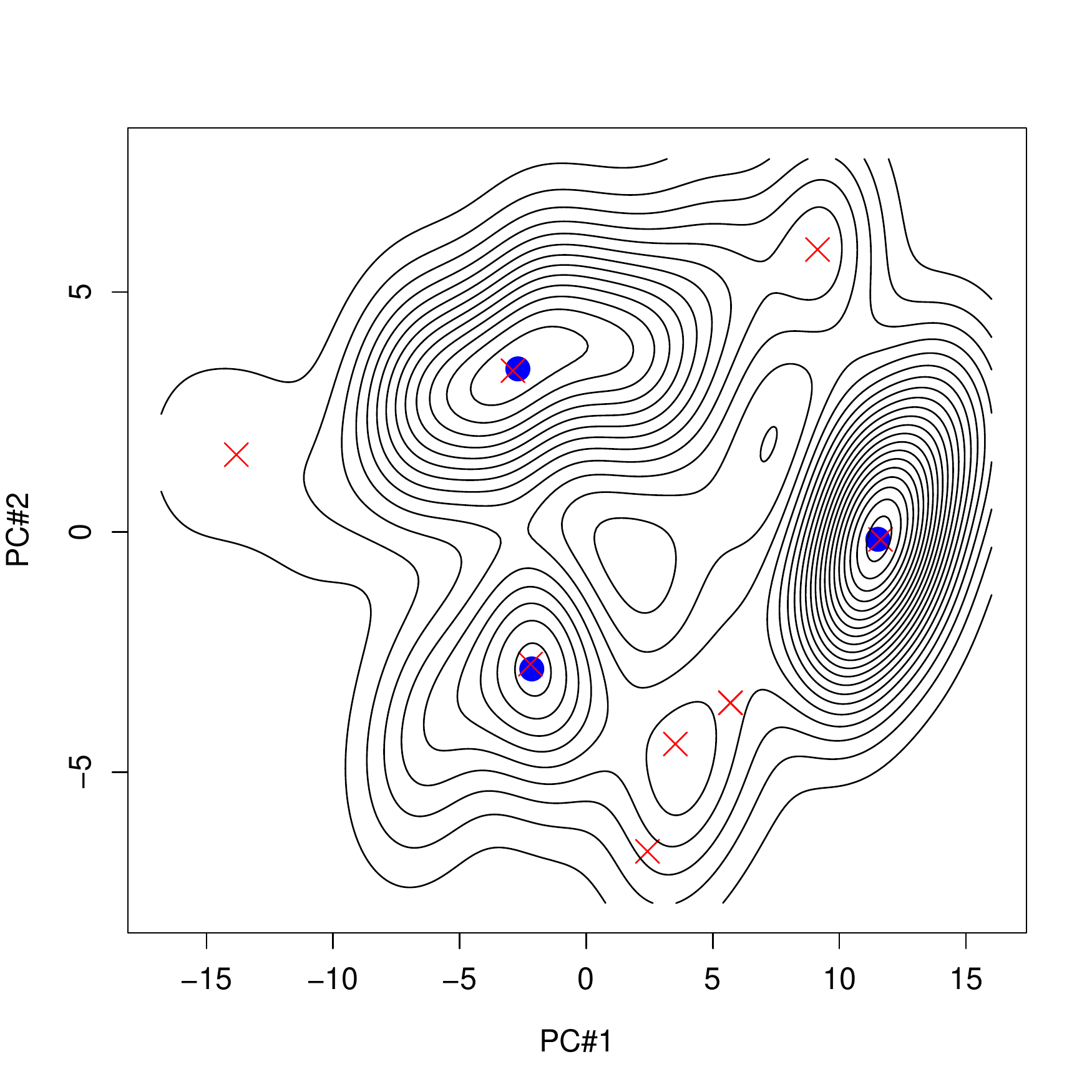}
\caption{Simplified example of the clustering performed by OCA to identify modes of density, shown in the space formed by the first two principal components.
The black lines are the contours of the data density. The blue circles and red crosses correspond to the 
modes of the large (0.025) and small (0.015) kernel sizes
respectively. At the lower resolution (blue circles), the three cluster
modes obtained correspond to clusters dominated by stars, quasars
and galaxies, although with significant contamination from other types.
\label{fig:oca}
}
\end{center}
\end{figure}

OCA has been tested on the semi-empirical SDSS star, galaxy and quasar libraries described in section~\ref{sect:training} for sources with a range of magnitudes. The BP/RP spectra were normalized to unit area, and only the first 15 principal components (PCs) in each of RP and BP were used, accounting for 99\% of the variance. We also included the first four moments of the BP and RP flux distributions, scaled to the range of values of the first PC. Figure~\ref{fig:oca} illustrates a simplified case of clustering for two kernel sizes and a reduced dataset.

\section{Validation, calibration, and in-mission development}\label{sect:validation}

The data processing phase of the mission comprises three vital tasks beyond applying Apsis to the Gaia data. These are validation, calibration, and in-mission software development. (The subsequent task of catalogue production is not covered in this article.) These we now discuss in turn.

Apsis will produce an enormous set of AP estimates on many different types of objects. A critical assessment of these results is an important part of the data processing. We refer to this as validation, and it will take place in two ways.  First, an internal validation examines (for example) the distributions of estimated parameters, their uncertainties, and correlations between them, and whether these agree with our expectations. For example, do we get an inordinately large number of low metallicity stars, or do we find unexpected correlations, such as \teff\ increasing with \a0?\footnote{There is a known, strong degeneracy in the spectra between \teff\ and \a0\ for individual stars (e.g.\ \citealp{2010MNRAS.403...96B,2012MNRAS.426.2463L}) which we attempt to account for, but is not of physical origin.}  
Such analyses may allow us to identify problems and thus improve the training data and algorithms. 
The internal validation will also compare the AP estimates for common objects between the different modules in Apsis.
Second, an external validation compares our AP estimates with external AP estimates, either for individual objects or for populations of objects. An example of the latter is to construct the HRD of known clusters, or compare metallicity or redshift distributions with published estimates.  While we would not automatically take non-Gaia estimates as being true, systematic differences between our and non-Gaia estimates may be indicative of problems.

The Apsis algorithms have been developed over the past years using simulated Gaia data.  The real Gaia data will of course differ from these. In particular, the response function and noise properties of the detectors may differ from expectations, and these will anyway evolve in unpredictable ways during the mission due to the progressive radiation damage of the CCDs.  Upstream data processing tasks may also need to change the way they process the data, producing data with different properties. The result is that the shape and noise properties of the spectra are likely to deviate from our current simulations.  Furthermore, the Apsis algorithms make extensive use of synthetic spectra for training. These differ from real spectra because of the approximations involved in modelling astrophysical sources.\footnote{An example mismatch affecting RVS is incorrect broadening parameters and assumptions, which affects in particular the Paschen lines of stars with \teff\ around 10\,000\,K \citep{1996MNRAS.279...25F}. In BP/RP, the largest effect is expected to arise for the cooler stars due to incomplete and poor molecular data, as shown for example by \citet{2011JPhCS.328a2005P} and \citet{2012A&A...547A.108L}.} These two issues -- spectral simulation and instrument simulation -- result in imperfectly modelled Gaia spectra, something we refer to as the ``spectral mismatch problem''.  As supervised algorithms depend on a match between their training data and the observed data, it is important that we accommodate these changes.  It is the goal of the calibration of the Apsis algorithms to correct for this.

A calibration can be achieved by applying corrections either to the training data before it is used, or to the APs produced by the estimation algorithm.  In the first approach, we use Gaia observations of labelled reference objects (i.e.\ with known APs) to modify the fluxes of the synthetic spectra, thereby producing a hybrid synthetic--real grid which is used for algorithm training.  One specific idea is to use the denser synthetic spectral grids to model the small scale variations of fluxes with APs, and the sparser observed spectral grids to model the larger scale variations \citep{LL:CBJ-044}.  This follows the assumption that synthetic spectra reproduce flux changes better than absolute fluxes.   The feasibility of this approach is under investigation.\footnote{Ideally we would just use Gaia observations of labelled objects as the complete training data sets and dispense entirely with modifying synthetic grids, but a sufficient set of labelled objects does not exist for this purpose.}  In the second approach, we instead model the AP deviations as a function of the main parameters. Although simpler, it is probably less accurate due to the loss of information from working with ``faulty'' spectra in the first place, so this is not being pursued.  Both approaches require that we obtain accurate APs by independent methods for a set of reference objects which Gaia will observe. This is being done explicitly for Gaia using ground-based higher resolution spectra, as discussed in section~\ref{sect:gbog}.

Another aspect of the calibration work is to improve the representation of the synthetic stellar spectra in the first place. Two routes are currently being followed to improve the spectra of FGK stars. The first is to move from classical 1D stellar atmospheres to 3D radiation-hydrodynamics simulations in order to better represent the effects of convection  \citep{2011JPhCS.328a2012C, 2013A&A...550A.103A, 2013arXiv1302.2621M}. Second, deviations from LTE will be taken into account by implementing the results of detailed statistical equilibrium calculations into the spectrum synthesis codes. This will be particularly important for modelling the calcium triplet lines dominating the RVS spectra \citep{2007A&A...461..261M, 2011MNRAS.418..863M}.

In addition to calibration and validation, we expect to have to adapt, during the data processing, how we use the Apsis algorithms. For example, we will inevitably have to modify our nominal strategy 
of which APs we attempt to estimate for which types of object at which magnitudes.
We may even find that we need to modify or change algorithms, or introduce new algorithms to deal with additional classes of object.  Indeed, we fully expect to have to modify and extend our spectral libraries to accommodate missing classes of objects or poorly modelled classes. We will also update the model of the Gaia instruments to match their in-flight properties as closely as possible, in order to produce more accurate training data sets. All of this will demand a continued software development and data simulation during the mission.

\subsection{High resolution spectral observations for Apsis stellar calibration}\label{sect:gbog}

To perform the calibrations described above, we need independent AP estimates of several thousand Gaia targets. 

A two-level procedure for those algorithms which estimate \teff, \logg, and \feh\ is foreseen.  At the first level, we define a set of benchmark stars made up of a small number of carefully selected, well-studied bright stars  (around 40 FGKM and 20 OBA stars). Their Hipparcos parallaxes, angular diameters and bolometric fluxes are known, and their masses have been determined in a homogeneous way, so their effective temperatures and surface gravities can be derived independently of spectroscopy. Reference metallicities for benchmark stars are determined from ground-based high-resolution spectra (see below) using several different methods. Details on the parameters and data for cool benchmark stars will be published in a series of forthcoming papers (and already in \citealp{2012A&A...547A.108L}). At the second level, we define a much larger set of several hundred reference stars covering the AP space more densely than the benchmark stars.  
Homogeneous APs for these stars are being determined from high-resolution spectroscopy and calibrated to the
benchmark stars. 

The necessary high-resolution spectra are being obtained in various observing programs. OBA stars have been observed with the HERMES spectrograph on the Mercator Telescope in Spain.  These will be supplemented with medium-resolution OBA (cluster) star spectra observed with VLT-Giraffe as part of the Gaia--ESO public spectroscopic survey. The FGKM stars have been observed with the NARVAL spectropolarimeter on the 2m Bernard Lyot Telescope at Pic du Midi in spectroscopic mode. High-resolution spectra of M dwarfs have been obtained in the infrared J-band with the CRIRES spectrograph \citep{2012A&A...542A..33O}. High quality spectra are also retrieved from the various public archives.  
So far the library comprises 79 spectra of 35 cool benchmark stars from NARVAL, UVES and HARPS observations, with resolutions greater than 70\,000 and SNRs greater than 200.
The benchmark star spectra will be published online in the {\sc SpectroWeb} database\footnote{http://spectra.freeshell.org/spectroweb.html} and the reference star spectra in the {\sc HHighRespect} database, currently under development.

\section{Outlook}\label{sect:outlook}

We have described the status of the Gaia classification system at the time of launch, prior to seeing any real data. During the course of the five year mission and the subsequent two or three years of processing before the final data release, this system will continue to evolve in light of the experience we gain with the data. Indeed, we anticipate substantial developments, which we will report in future publications.

Our classification approach involves a combination of supervised and unsupervised algorithms. The former are critically dependent on an accurate representation of the target sources, and our training data sets will need considerable optimization during the data processing phase. This will involve improvements to the simulations as well as the use of ground-based data to calibrate our training data.

The system developed so far makes some idealized assumptions about the data and the upstream spectral processing by the other CUs.  For example, the exact impact of radiation damage on the CCDs and therefore on the combined BP/RP and RVS spectra is hard to model. Being a slitless spectrograph, some BP/RP spectra will overlap (and likewise for RVS). Although this is accommodated in the spectral extraction, imperfect removal of overlap will leave systematic residuals.  
Analysing the real data will be an important learning experience. 

It should be appreciated that the main objective of Apsis is to provide reasonably accurate parameter estimates for a broad class of objects covering a large fraction of the catalogue. We do not aim to do everything possible. For example, while we try to identify white dwarfs, we do not (yet) attempt to estimate their parameters. We are likewise aware that almost any narrow class of objects could be given more targeted treatment, which may result in more accurate AP estimates through, for example, a more focused use of the data or by adopting different source models. A combination of Gaia data with non-Gaia data will be particularly beneficial for some classes of object, but this is beyond the remit of CU8. Some such work is planned by the DPAC within CU9 (responsible for the data releases), and the resulting hybrid catalogues would be published in the data releases.  We note finally that we hope to publish not only the results from Apsis, but also the software, to enable the scientific community to obtain their own AP estimates with their own training data sets, for instance. 
There is no natural divide between ``data processing'' and ``scientific analysis'', and we hope that in the course of exploiting the Gaia data the community will take up the challenge to extend and to improve our work.

\begin{acknowledgements}

  We acknowledge the contributions of our former colleagues who were involved in earlier stages of this work. We remember in particular the contributions of our prematurely deceased colleague Jo\"el Poels.  We are grateful to the members of the Gaia DPAC CU2 for the development of the GOG simulator.  Many of the GOG simulations used in this work were produced by CU2 using the supercomputer MareNostrum at the Barcelona Supercomputing Center -- Centro Nacional de Supercomputaci\'on.  Parts of this work have been funded by various funding agencies including: DLR (German space agency), grant 50 QG 1001; CNES (French space agency); ASI, under contract I/058/10/0 (Gaia Mission - The Italian Participation in DPAC); Spanish Ministry of Science and Innovation, through grants AyA2009-14648-C02-02 and AyA2012-39551-C02-02; Spanish Ministry of Economy, through grants AyA2008-02156 and AyA2011-24052; Swedish National Space Board; ESA PRODEX programme ExSciGaia (contract 4000106135); Belgian Federal Science Policy Office in connection with the ESA PRODEX programmes `Gaia-DPAC QSOs' and `Binaries, extreme stars and solar system objects' (contract C90290). SIGAMM computing centre of the Observatoire de la C\^ote d'Azur.

\end{acknowledgements}

\bibliographystyle{aa}
\bibliography{apsis2013}

\end{document}